\documentclass[fleqn,usenatbib]{mnras}

\usepackage{newtxtext,newtxmath}

\usepackage[T1]{fontenc}
\usepackage{ae,aecompl}

\usepackage{graphicx}	

\newcommand{\sfr}{\mathrm{SFR}}

\newcommand{\Ha}{H$\alpha$}

\newcommand{\OIII}{[\ion{O}{III}]}
\newcommand{\NII}{[\ion{N}{II}]}
\newcommand{\FeII}{[\ion{Fe}{II}]}

\newcommand{\SIII}{[\ion{S}{III}]}

\newcommand{\phistar}{\Phi^{\star}}
\newcommand{\Lstar}{L^{\star}}

\title[Confirmation of Two  $z = 2.24$ MAMMOTH overdensities]{MAMMOTH: Confirmation of Two Massive Galaxy Overdensities at  $z = 2.24$ with H$\alpha$ Emitters}

\author[X.~Z. Zheng et al.]{
Xian~Zhong Zheng,$^{1,2}$\thanks{E-mail: xzzheng@pmo.ac.cn (XZZ); zcai@tsinghua.edu.cn (ZC)}
Zheng Cai,$^{3\color{blue}\star}$
Fang~Xia An,$^{4}$
Xiaohui Fan,$^{5}$
Dong~Dong Shi$^{1,2,5}$
\\
$^{1}$Purple Mountain Observatory, Chinese Academy of Sciences, 10 Yuanhua Road, Nanjing 210023, China\\
$^{2}$School of Astronomy and Space Sciences, University of Science and Technology of China, Hefei 230026, China \\
$^{3}$Department of Astronomy, Tsinghua University,  Beijing 100084, China, \\
$^{4}$Inter-University Institute for Data Intensive Astronomy, and Department of Physics and Astronomy, University of the Western Cape, \\  Robert Sobukwe Road, Bellville 7535, Cape Town, South Africa \\
$^{5}$Steward Observatory, University of Arizona, 933 N. Cherry Ave., Tucson, AZ 85721, USA
}

\date{Accepted 2020 September 16. Received 2020 September 14; in original form 2019 December 10}

\pubyear{2020}

\begin{document}
\label{firstpage}
\pagerange{\pageref{firstpage}--\pageref{lastpage}}
\maketitle

\begin{abstract}
Massive galaxy overdensities at the peak epoch of cosmic star formation provide ideal testbeds for the formation theories of galaxies and large-scale structure. We report the confirmation of two massive galaxy overdensities at  $z = 2.24$, BOSS1244 and BOSS1542, selected from the MAMMOTH project using Ly$\alpha$ absorption from the intergalactic medium over the scales of 15$-$30\,$h^{-1}$\,Mpc imprinted on the quasar spectra.  We use \Ha\ emitters (HAEs) as the density tracer and identify them using deep narrowband $H_2S1$ and broadband $K_{\rm s}$ imaging data obtained with CFHT/WIRCam. In total, 244 and 223 line emitters are detected in these two fields, and $196\pm 2$ and $175\pm 2$ are expected to be HAEs with an \Ha\ flux of $> 2.5\times 10^{-17}$\,erg\,s$^{-1}$\,cm$^{-2}$ (corresponding to an SFR of $>$5\,M$_\odot$\,yr$^{-1}$).   The detection rate of HAE candidates suggests an overdensity factor of $\delta_{\rm gal}=5.6\pm 0.3$ and $4.9\pm 0.3$  over the volume of $54\times32\times32$\,cMpc$^3$. The overdensity factor  increases  $2-3$ times when focusing on the high-density regions of scales $10-15$\,cMpc. Interestingly,  the HAE density maps reveal that  BOSS1244 contains a dominant structure,  while BOSS1542 manifests as a giant filamentary structure.  We measure the \Ha\ luminosity functions (HLF), finding that BOSS1244's HLF is nearly identical to that of the general field at the same epoch, while BOSS1542 shows an excess of HAEs with high \Ha\ luminosity,  indicating the presence of enhanced star formation or AGN activity. We conclude that the two massive MAMMOTH overdensities are undergoing a rapid galaxy mass assembly. 
\end{abstract}




\section{Introduction}

Understanding the formation of galaxy clusters is a central task in modern astrophysics \citep{Berrier2009,Allen2011}.  While the standard $\Lambda$CDM model is successful at reproducing the dark matter-driven perspective of cluster formation (e.g., the abundance and clustering properties), the physical processes that regulate the mass assembly of cluster member galaxies and influence the baryons within a cluster through feedback remain to be fully understood \citep{Kravtsov2012,Schaye2015}.  Compared with the general field, galaxy clusters contain more massive galaxies and amplify details of these baryonic processes, including gas cooling, star formation, stellar feedback, black hole activity,  galaxy merging and environmental effects, thus making them unique testbeds for theoretical models of galaxy formation \citep{Overzier2016}.

It has long been known that the dense environment of galaxy clusters dramatically affect galaxy properties. The massive early-type galaxies in the clusters tend to form at earlier epochs, indicating that their progenitors would  be actively star-forming galaxies (SFGs) in galaxy protoclusters at $z\ga 2-3$ \citep{Thomas2005}.  Indeed, cluster galaxies have lower star formation rates (SFRs) than field galaxies in the local universe \citep[e.g.,][]{Dressler1984, Kauffmann2004, Blanton2009,von2010, Owers2019},  while this trend is found to be reversed at $z>1$ \citep{Elbaz2007,Tanaka2010, Koyama2013, Dannerbauer2014, Tran2015, Umehata2015, Hayashi2016, Shimakawa2018a}.  A higher fraction of active galactic nuclei (AGNs) was reported in some $z\sim1-3$ protoclusters compared with the general field at the same epoch, indicating an enhanced growth of supermassive black holes (SMBHs)  in the high-density environment \citep{Lehmer2009,Digby-North2010, Martini2013,Krishnan2017}. Similarly, the fraction of galaxy mergers \citep{Hine2016,Watson2019} and galaxy gas fraction \citep{Noble2017,Coogan2018} are likely to be higher in $z>2$ protoclusters, although the fraction of massive gas-rich SFGs in the central regions of protoclusters depends on their evolutionary stage \citep{Casey2015,Wang2018,Shimakawa2018a, Zavala2019}. Nevertheless, how these distant SFGs  evolve into the local massive galaxies in different cluster environments is still not yet clear \citep[e.g.,][]{DeLucia2007,Lidman2012, Contini2016,Casey2016,Shimakawa2018b}. In particular, where and how different environmental interactions play roles in shaping galaxy properties remain open questions.    
Galaxy protoclusters at the peak epoch of cosmic star formation and black hole growth ($z\sim 2-3$; \citealt{Madau2014}) provide a useful probe of the rapid mass assembly of galaxies in relation to structure formation \citep[][]{Bond1996,Boylan-Kolchin2009,Brodwin2013,Chiang2017}.  Investigating massive protoclusters and the properties of their member galaxies at this peak epoch will provide key constraints on the  environmental dependence of the galaxy evolution and black hole growth.

A protocluster refers to an unviralized structure of all the dark matter and baryons that will assemble into a present-day galaxy cluster.  Galaxy protoclusters at $z>2$ are expected to have an average overdensity of $\rho/\bar\rho \approx 2$ over a scale of $\ga 20\,h^{-1}$\,co-moving\,Mpc (cMpc) \citep{Muldrew2015,Lovell2018}.  In practice, one can identify galaxy overdensities of a given scale at high $z$ but whether they are protoclusters depends on the scale and their surrounding gravitational environments. Generally, massive overdensities over large scales of $\ga 10-30\,h^{-1}$\,cMpc are naturally represent protoclusters while small-scale overdensities may be either the progenitors of local groups or part of the protoclusters. 
Yet, a number of $z>2$ protoclusters have been spectroscopically identified. However, few of them were initially identified as massive overdensities at a scale of $\ga 20\,h^{-1}$\,cMpc. These protoclusters were selected by various means and thus often biased by selection effects \citep[e.g.,][]{Shi2019}.
Deep cosmic surveys are used to detect protoclusters at high $z$ \citep[e.g.,][]{Lemaux2014, Cucciati2014, Yuan2014, Tran2015, Chiang2015, Wang2016, Toshikawa2016}.   Rare massive sources, e.g., quasars or bright radio galaxies, usually reside in dense environments and can also be used as protocluster indicators \citep{Venemans2007,Hayashi2012,Onoue2018}.  Surveys for galaxy clusters relying on either the Sunyaev-Zel'dovich (SZ) effects \citep{Bleem2015} or excess of red-sequence galaxies \citep{Gilbank2011, Strazzullo2016} are biased to pick up relaxed ones mostly at $z<1.5$, containing hot gas and/or a large fraction of quenched massive galaxies.  The sample of confirmed protoclusters at  $z=2-4$ selected by these approaches are incomplete and difficult for statistical comparisons with hierarchical models of structure formation \citep{Chiang2013}.  Moreover, the evolution of the most massive haloes at high $z$ are essentially determined by the surrounding  density field on large scales of $\ga 10\,h^{-1}$\,cMpc \citep{Angulo2012}.  The identified protoclusters at small scales might not necessarily evolve into the present-day massive clusters.

Ly$\alpha$ forest optical depth is predicted to be strongly correlated with dark matter overdensity at scales of $\gtrsim 3\,h^{-1}$\,cMpc and the correlation peaks at $10-30\,h^{-1}$\,cMpc \citep[e.g.,][]{Kollmeier2003}.  \citet{Cai2016} demonstrated with simulations that the intergalactic medium (IGM) traces the underlying dark matter density field, and the strongest IGM Ly$\alpha$ absorptions mostly trace massive overdensities at the scale of 15\,$h^{-1}$\,cMpc.  
Based on this correlation,  a novel approach (MAMMOTH: Mapping the Most Massive Overdensities Through Hydrogen) has been developed for identifying such mass/galaxy overdensities at  $z=2-3$, traced by groups of Coherently Strong Ly$\alpha$\ Absorption (CoSLA) imprinted on the spectra of a number of background quasars \citep{Cai2016}. This method is inherently less biased than many other techniques because the H\,{\scriptsize I} density is closely correlated with matter density over large scales.  It also covers a much larger survey volume, when using the large quasar absorption line database from spectroscopic surveys such as SDSS and BOSS. 
This technique has been successfully confirmed with the discovery of the  BOSS1441 protocluster at  $z=2.32$ using the early data release of SDSS-III \citep{Cai2017}.  The spectroscopic database from SDSS-III allow us to search for more massive overdensities of scales of $10-30\,h^{-1}$\,cMpc over dramatically larger volumes.

We aim to construct a statistical sample of MAMMOTH overdensities and fully quantify and characterize their member galaxies. We use a pre-existing narrowband filter $H_2S1$ to detect HAEs at  $z=2.24$,  which resulted in the selection of  two  $z=2.24$ overdensities traced by extreme groups of IGM Ly$\alpha$ absorption systems from  SDSS-III quasar spectra.  In this work, we present  the results of confirmation of the two massive overdensities with \Ha\ emitters. A detailed analysis of member \Ha\ emission-line galaxies will be presented in a subsequent paper (Shi.~D.~D. et al. in prep). The selection of a sample of MAMMOTH overdensities and implications to the formation of cosmic structures will be given in Cai~Z. et al. (in prep). 

In Section~\ref{sec:selection}, we introduce how the two targets are selected. Section~\ref{sec:obs} presents the near-infrared imaging observations and data reduction.  Our results are given in Section~\ref{sec:results}. We discuss and summarize our results in Section~\ref{sec:sum}. A standard $\Lambda$CDM cosmology with $H_0$=70\,km$^{-1}$\,Mpc$^{-1}$, $\Omega _{\rm \Lambda}$=0.7 and $\Omega _{\rm m}$=0.3  and a \citet{Kroupa2001} Initial Mass Function (IMF) are adopted throughout the paper. All magnitudes are referred to the AB system unless mentioned otherwise.

\section{Selection of two   $z=2.24$ MAMMOTH targets} \label{sec:selection}

Our goal is to confirm the massive overdensity candidates from MAMMOTH using \Ha\ emission-line objects at $z=2.246\pm 0.021$ selected from narrow-band $H_2S1$ ($\lambda_{\rm c}=2.130\,\micron$, $\Delta\lambda=0.0293\,\micron$) and broad-band $K_{\rm s}$ filters on CFHT/WIRCam. 
The MAMMOTH overdensities are selected using the IGM Ly$\alpha$ forest absorption systems from the SDSS-III \citep{Alam2015} over a sky coverage of 10,000\,deg$^2$.  To match the $H_2S1$ filter,  only the deep IGM absorption with the redshift range of $z=2.246\pm 0.021$ are used.

\begin{figure}
\centering
\includegraphics[width=0.49\columnwidth]{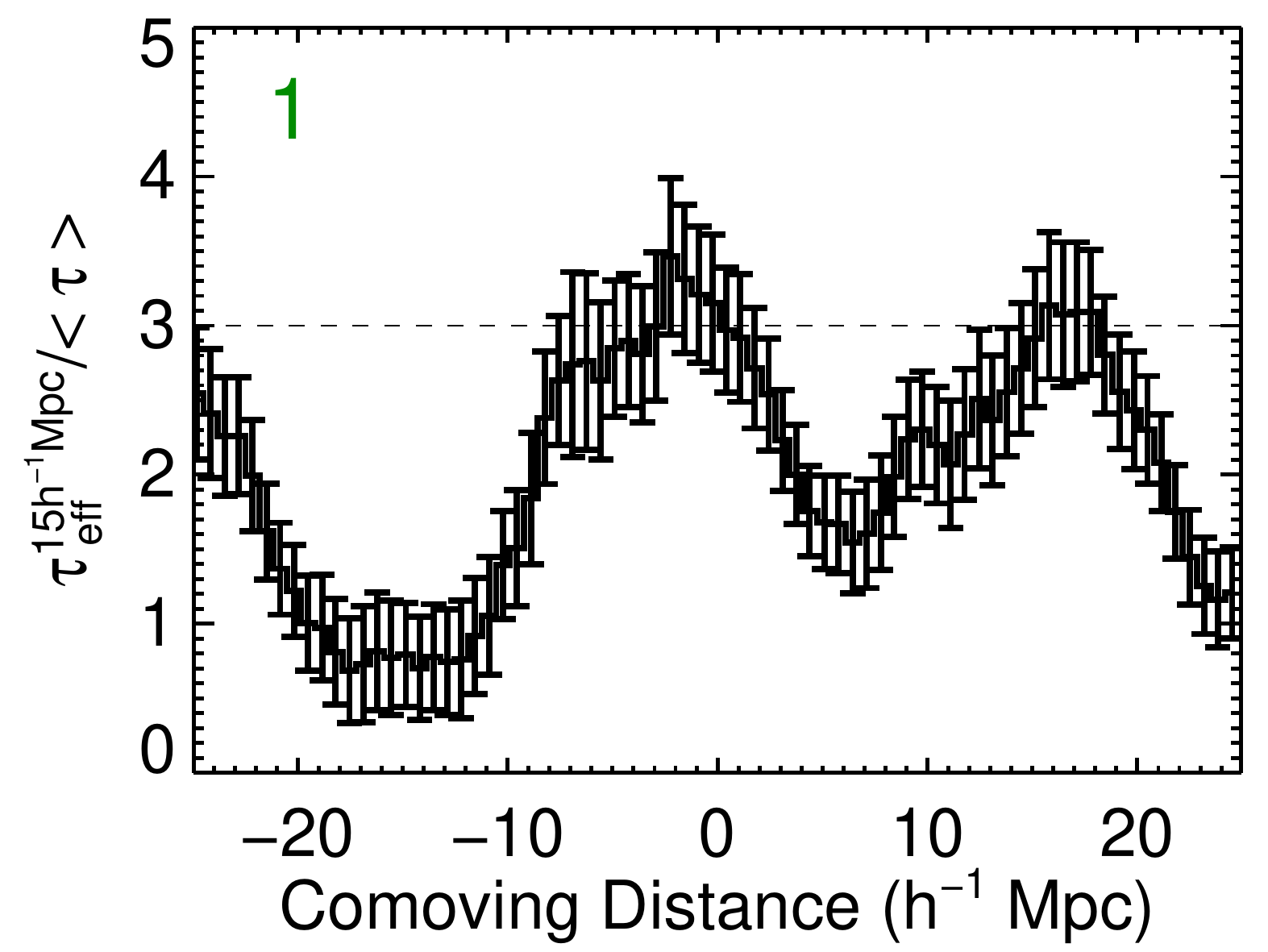}
\includegraphics[width=0.49\columnwidth]{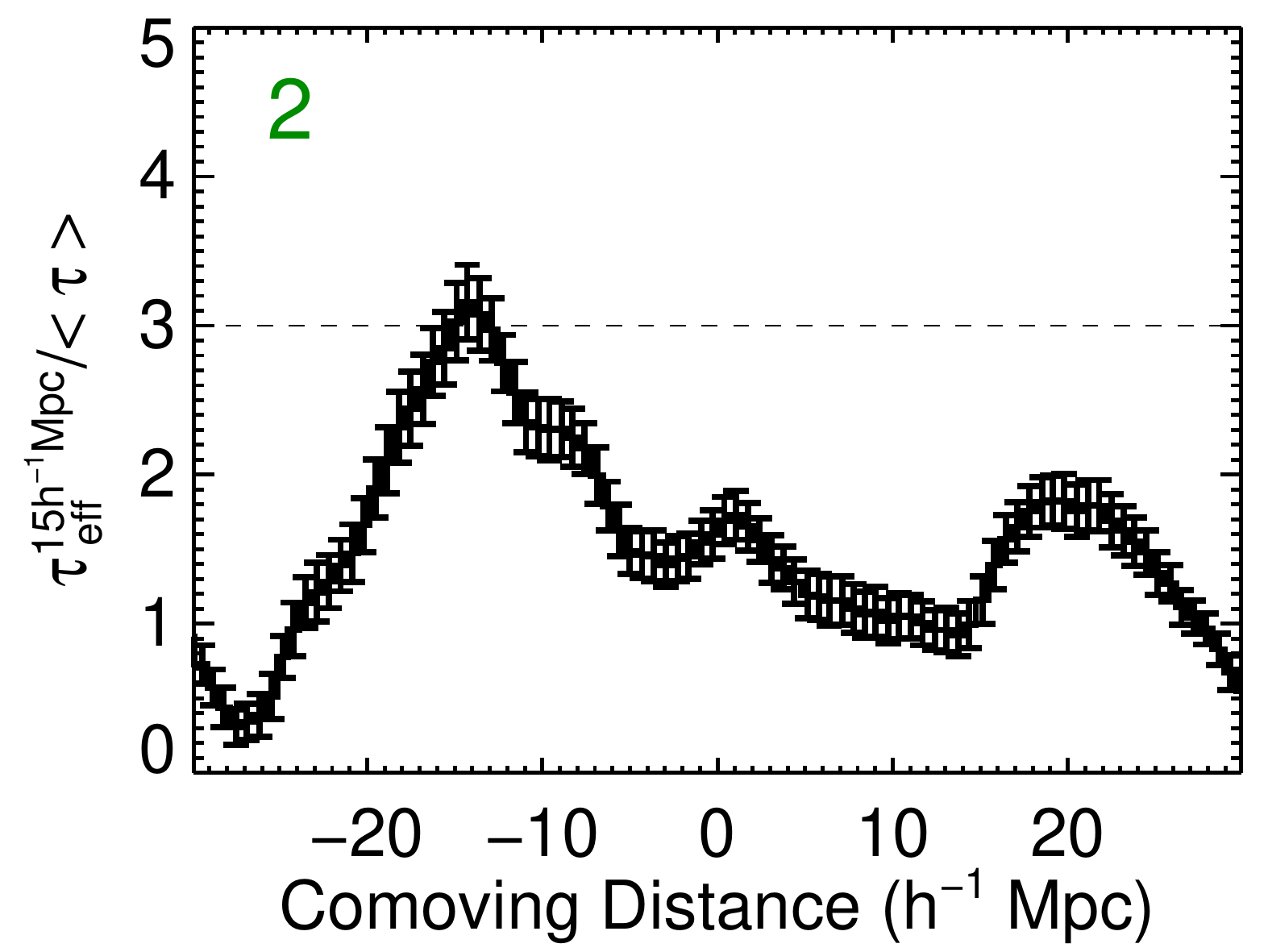}
\includegraphics[width=0.49\columnwidth]{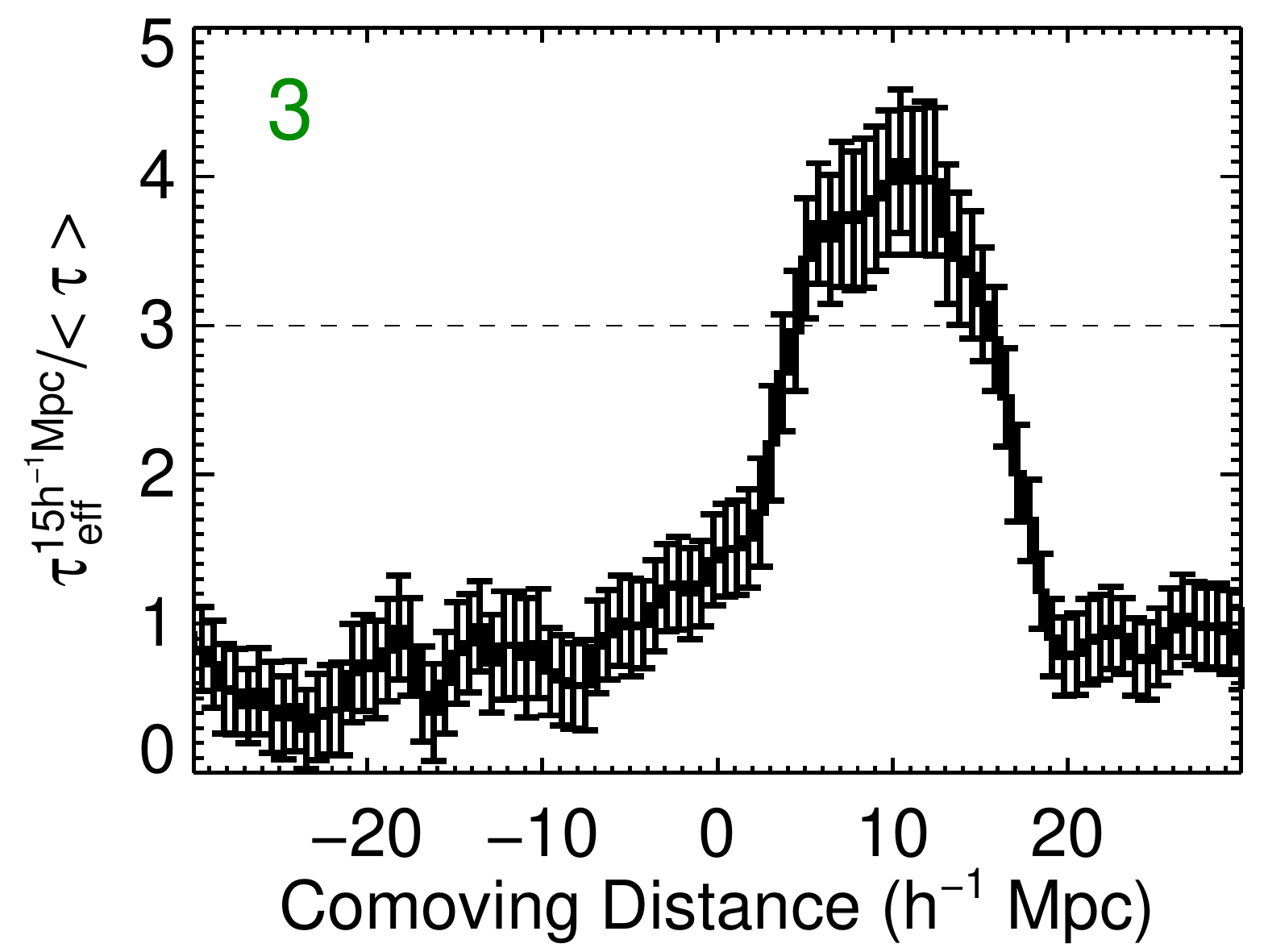}
\includegraphics[width=0.49\columnwidth]{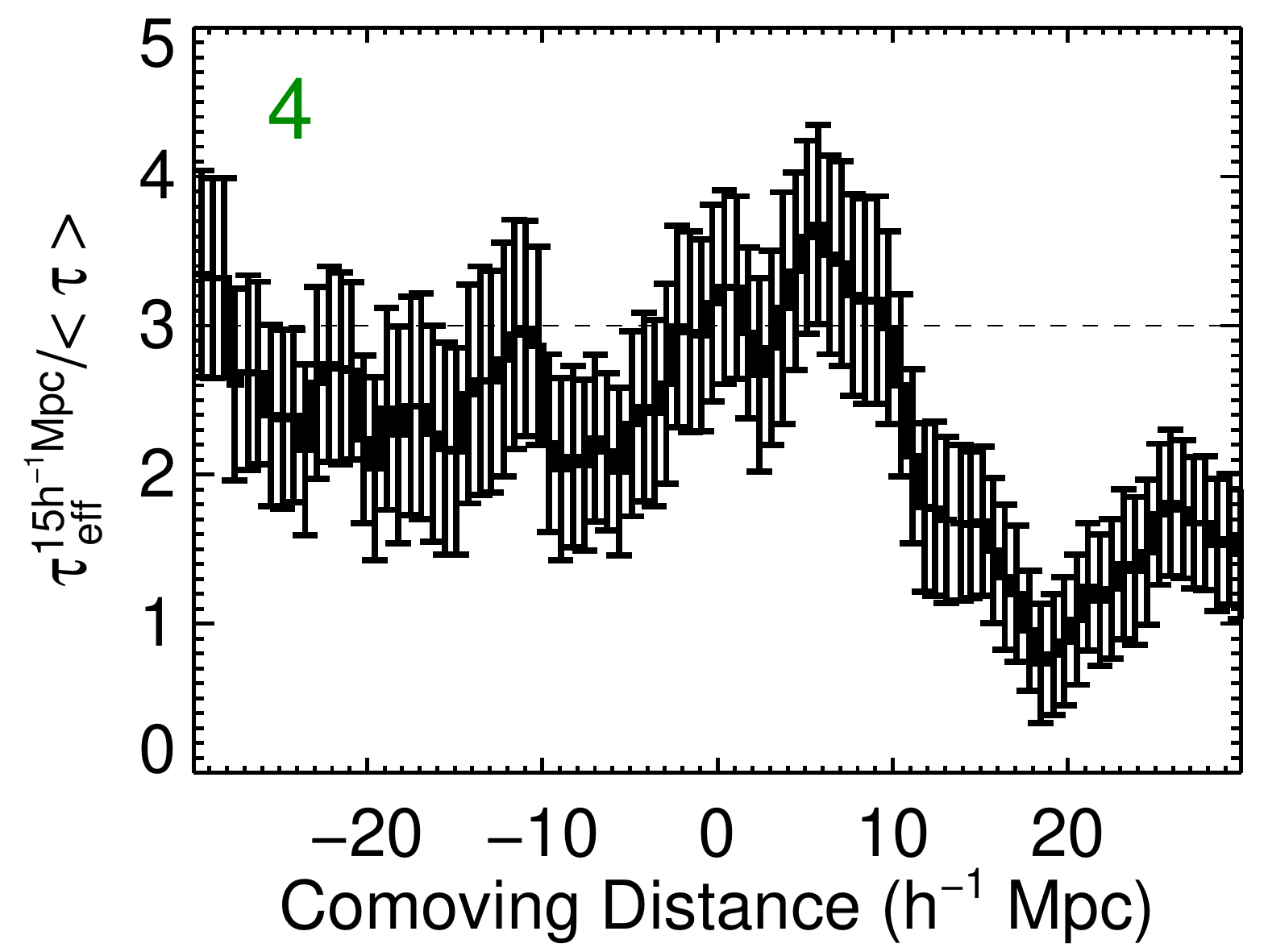}
\caption{Effective optical depth ($\tau_{\rm eff}$) derived from Ly$\alpha$ absorption lines at $z\sim 2.24$ imprinted on the spectra of four quasars in the BOSS1244 field.}\label{fig:abs1}
\end{figure}

\begin{figure}
\centering
\includegraphics[width=0.49\columnwidth]{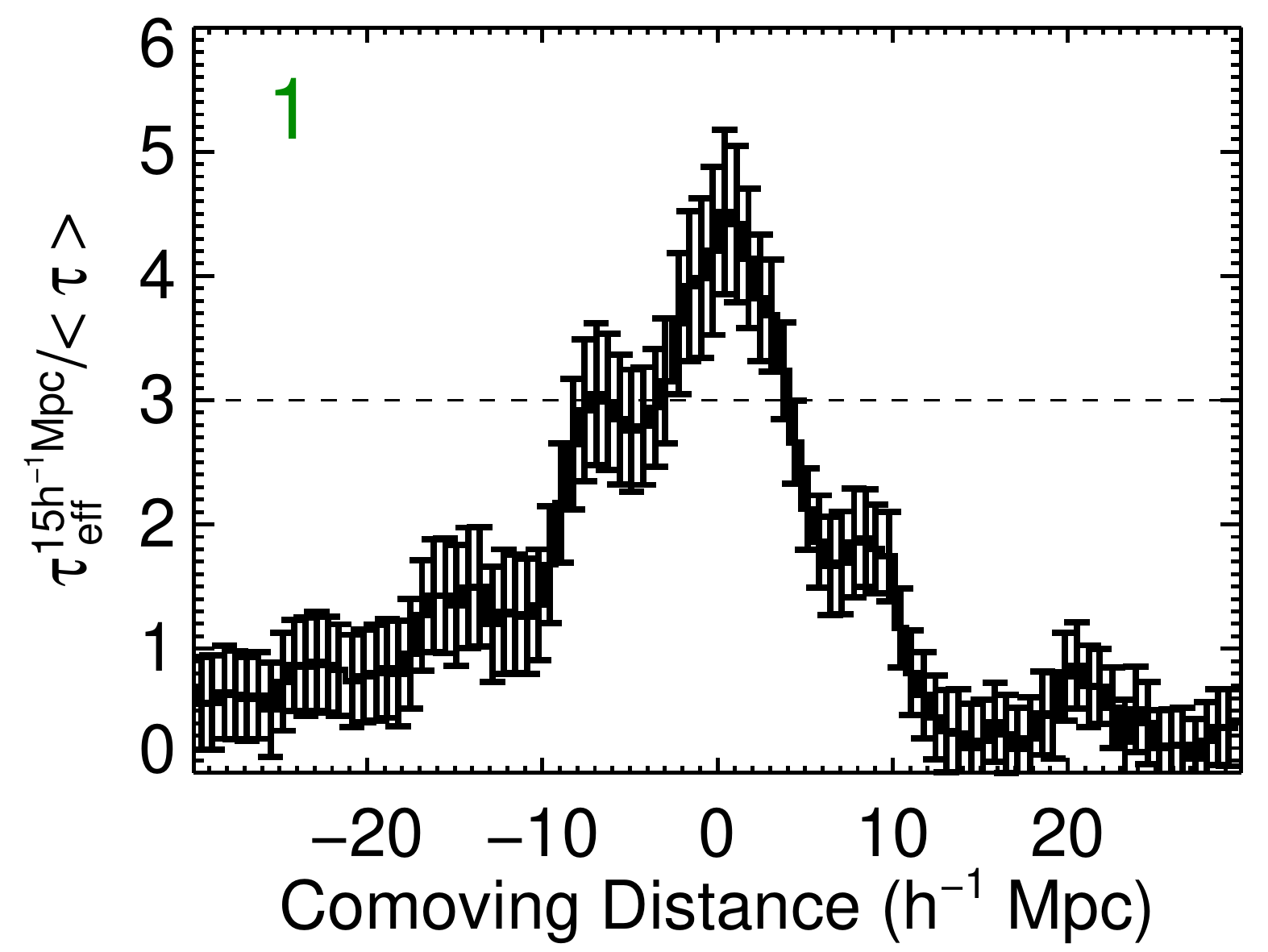}
\includegraphics[width=0.49\columnwidth]{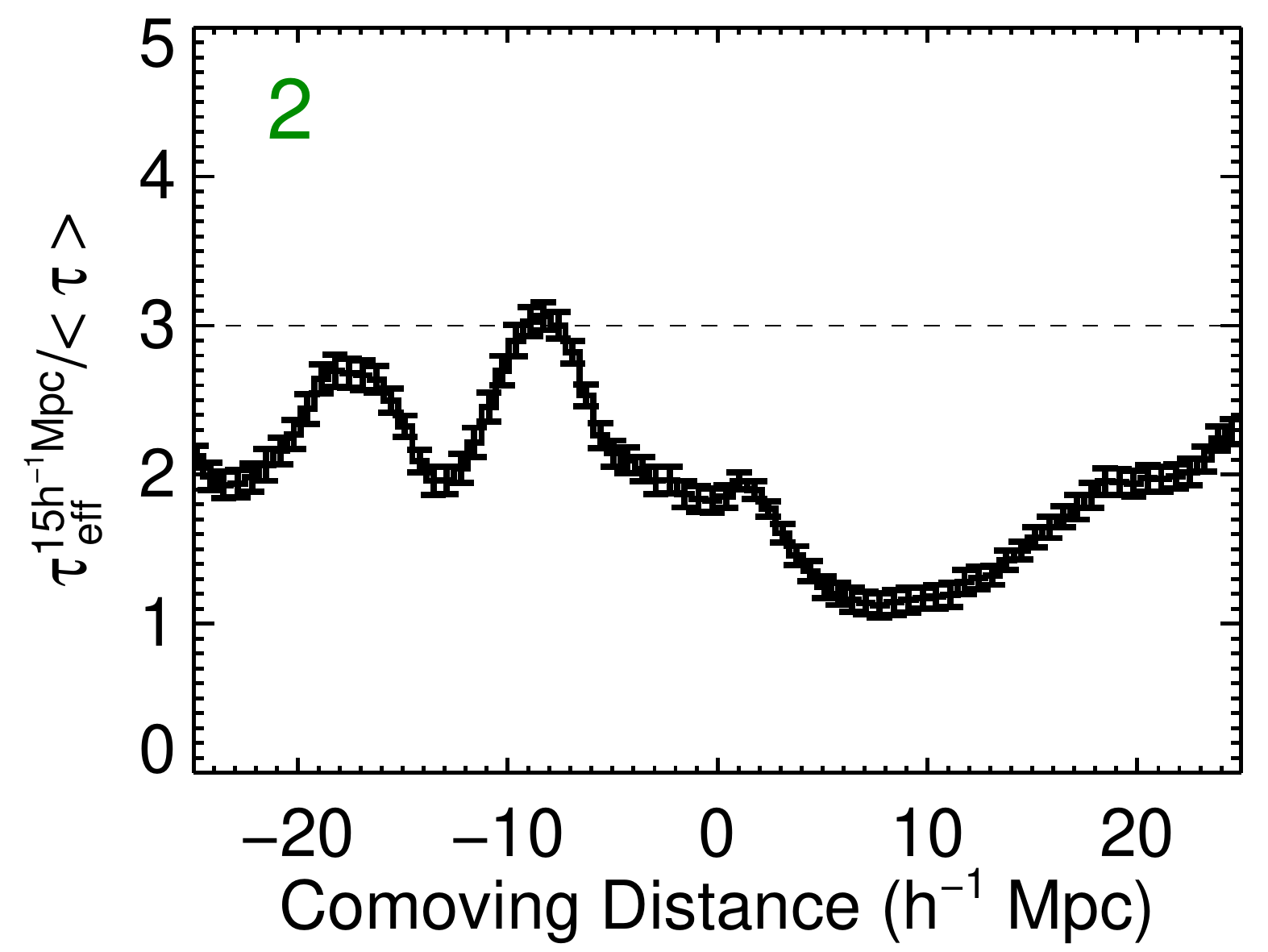}
\includegraphics[width=0.49\columnwidth]{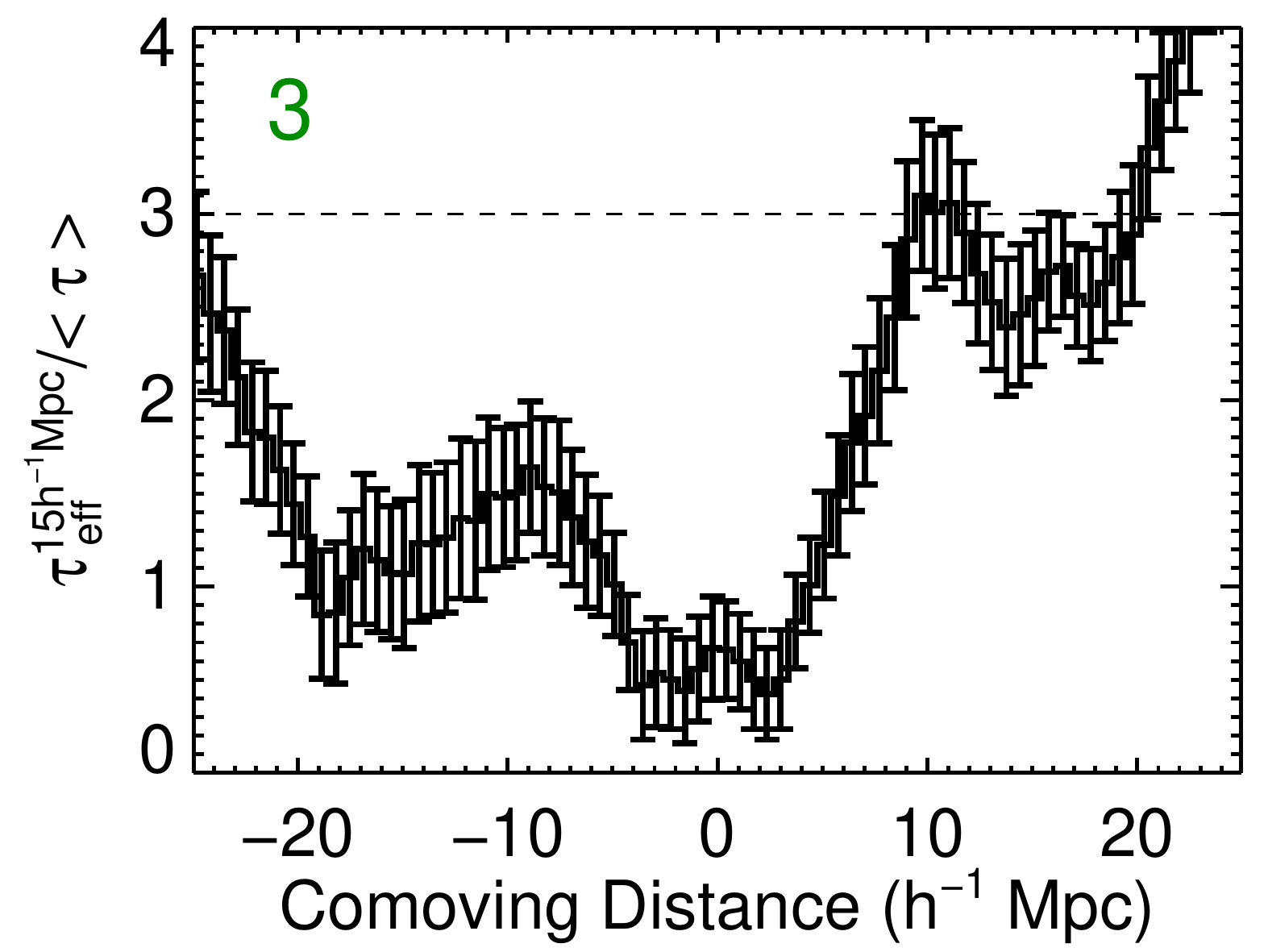}
\includegraphics[width=0.49\columnwidth]{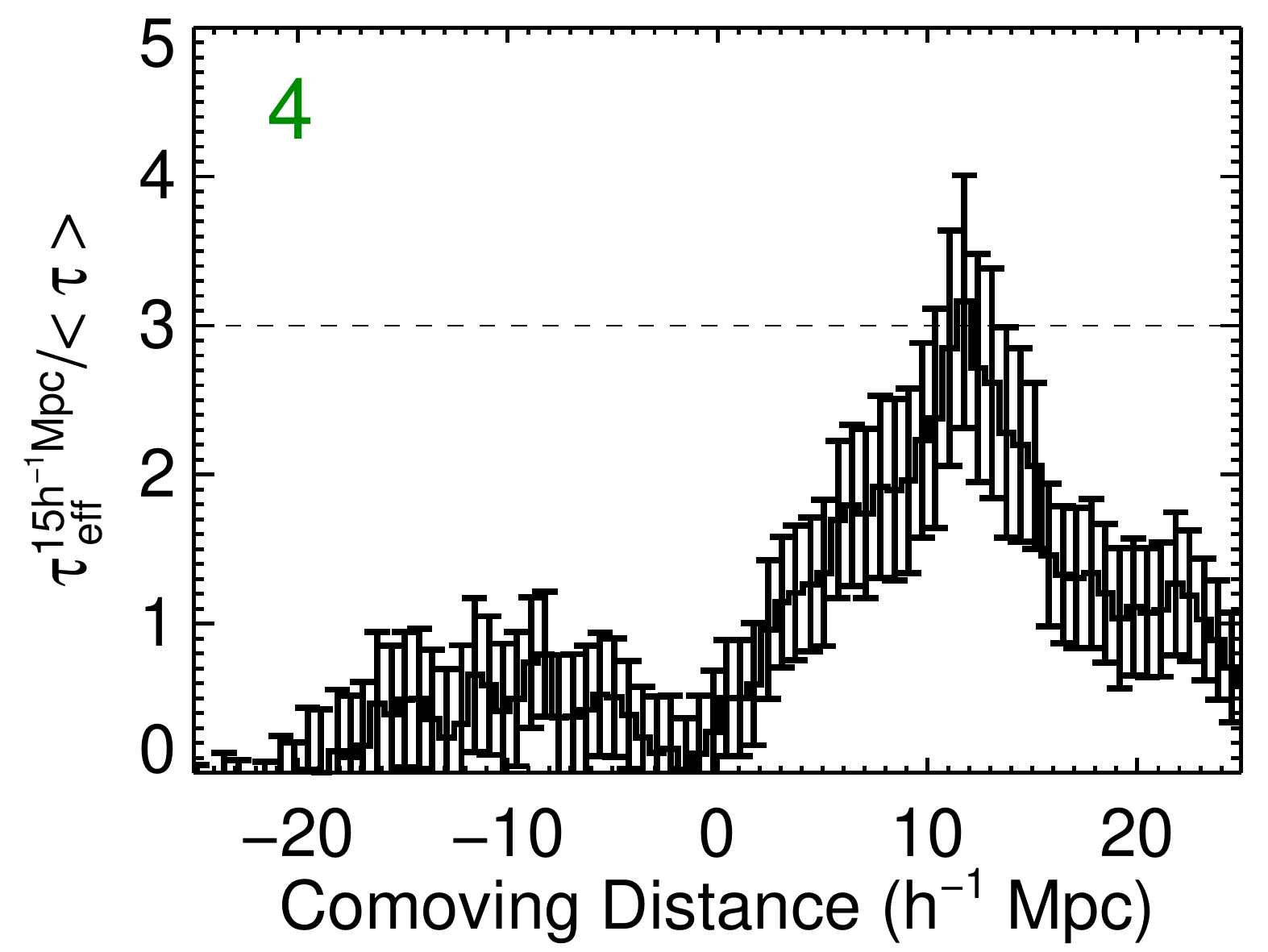}
\caption{Effective optical depth ($\tau_{\rm eff}$) derived from Ly$\alpha$\ absorption lines at $z\sim 2.24$ imprinted on the spectra of four quasars in the BOSS1542 field.}\label{fig:abs2}
\end{figure}

Following \citet{Cai2016}, the deep Ly$\alpha$ absorbers are selected by selecting regions where the effective optical depth ($\tau_{\rm eff}$) over 15\,$h^{-1}$\,Mpc ($=15$\AA)  is 4$\times$ higher than the mean optical depth at $z=2.2$.  Using the selection criteria described in detail in Cai et al. (in prep), we removed the contaminant DLAs which also causing large EW absorption based on the Ly$\alpha$ absorption profiles.  We then select the fields with the highest density of deep IGM absorption. From the complete SDSS-III quasar database, we identified two target fields, BOSS1244 and BOSS1542, suitable for observing in the Spring-Summer semester. The two fields have  groups of IGM strong absorption  systems comparable to those in the BOSS1441 field \citep{Cai2017} and also contain  several quasi-stellar objects (QSOs; i.e., quasars) at the same redshift.  Figure~\ref{fig:abs1} and Figure~\ref{fig:abs2} present the effective optical depth $\tau_{\rm eff}$  along the line of sight derived from strong Ly$\alpha$ absorption lines by absorbers at $z\sim2.24$  imprinted on quasar spectra in the two selected fields. These absorbers probed by background quasars  spread over a scale of 15\,$h^{-1}$\,Mpc.

\section{Observations and data reduction}\label{sec:obs}

We used WIRCam on board the Canada-France-Hawaii Telescope (CFHT)  to obtain deep near-infrared (NIR) imaging of the two MAMMOTH fields in both the narrow $H_2S1$ ($\lambda_{\rm c}=2.130\,\micron$, $\Delta\lambda=0.0293\,\micron$) and broad $K_{\rm s}$ ($\lambda_{\rm c}=2.146\,\micron$, $\Delta\lambda=0.3250\,\micron$) filters (PI: FX~An). The observations were carried out with the regular QSO mode under a median seeing  of $0\farcs65 - 0\farcs8$.  WIRCam has a field of view of $20\arcmin\times 20\arcmin$, covered by four 2048$\times$2048 HAWAII2-RG detectors with a pixel scale of 0$\farcs$3\, pixel$^{-1}$. The gaps between detectors are 45$\arcsec$. The observations were dithered to cover gaps between detectors and correct for bad pixels. We centered the FOV of WIRCam at the centers of BOSS1244 (R.A.=12:43:55.49, Dec.=+35:59:37.4) and BOSS1542 (R.A.=15:42:19.24, Dec.=+38:54:14.1) for the epoch of J2000.0. 
The total integration times are 7.18 and 4.96\,hours for the $H_2S1$ and  $K_{\rm s}$ observations in BOSS1244, and 7.275 and 5.17\,hours for the $H_2S1$ and  $K_{\rm s}$ observations in BOSS1542, respectively.  Each exposure takes 190\,s for the $H_2S1$ filter (194\,s in BOSS1542) and 20\,seconds for the $K_{\rm s}$ filter. Accounting for the overall overhead time (10\,s per exposure), the total observing time is 7.50\,hours for each of the two bands in BOSS1244, and 7.65\,hours  for $H_2S1$ and 7.75\,hours for $K_{\rm s}$ in BOSS1542. In total 30.40\,hours of telescope time were used in our observing program of two MAMMOTH fields.

The data reduction was carried out following \citet{An2014}. 
The reduced $H_2S1$ and $K_{\rm s}$ images were calibrated in astrometry using compact sources from SDSS. In total $\sim$700 SDSS compact sources with $12.0<z[mag]<20.5$ in the BOSS1244 field and 1,985 compact sources with $12.0<z[mag]<20.5$ in the BOSS1542 field are used for astrometric calibration, giving an astrometric accuracy of $\sim 0\farcs$1. Co-adding 136/893 and 135/930 frame $H_2S1$/$K_{\rm s}$ science images produced the final science images and the exposure time maps in BOSS1244 and BOSS1542, respectively. The points sources from 2MASS catalog are used to perform photometric calibration. 
In total 186 and 283 point sources with $12.6<K_{\rm s}[mag]<15.5$ in the two fields are selected for photometric calibration. An empirical point spread function (PSF) is built from these stars and used to derive aperture correction.  The photometric calibration reaches an accuracy of 1\%  for the selected stars in our mosaic $H_2S1$ and $K_{\rm s}$ images.

\begin{figure}
\centering
\includegraphics[width=0.48\columnwidth]{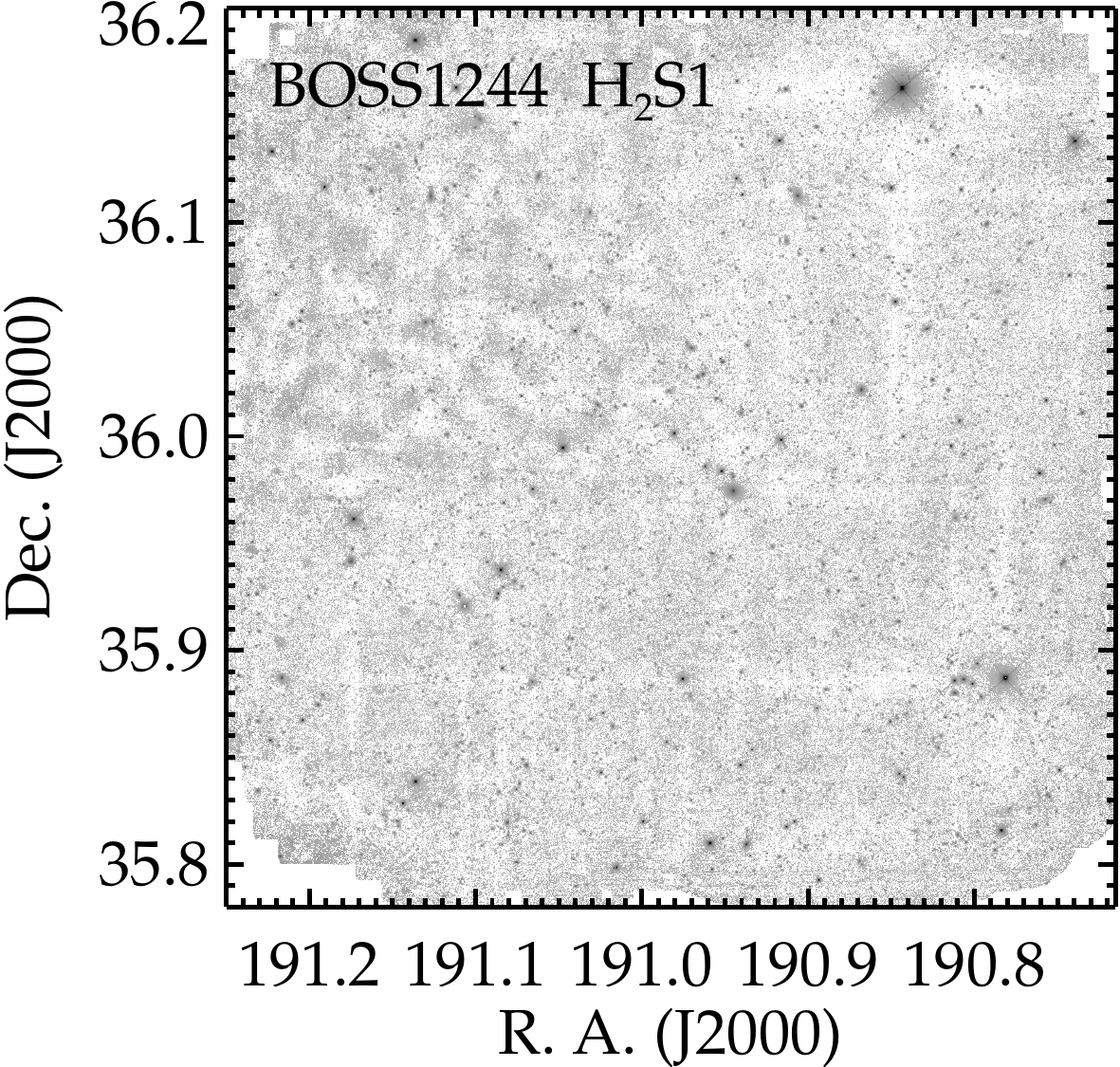}
\includegraphics[width=0.48\columnwidth]{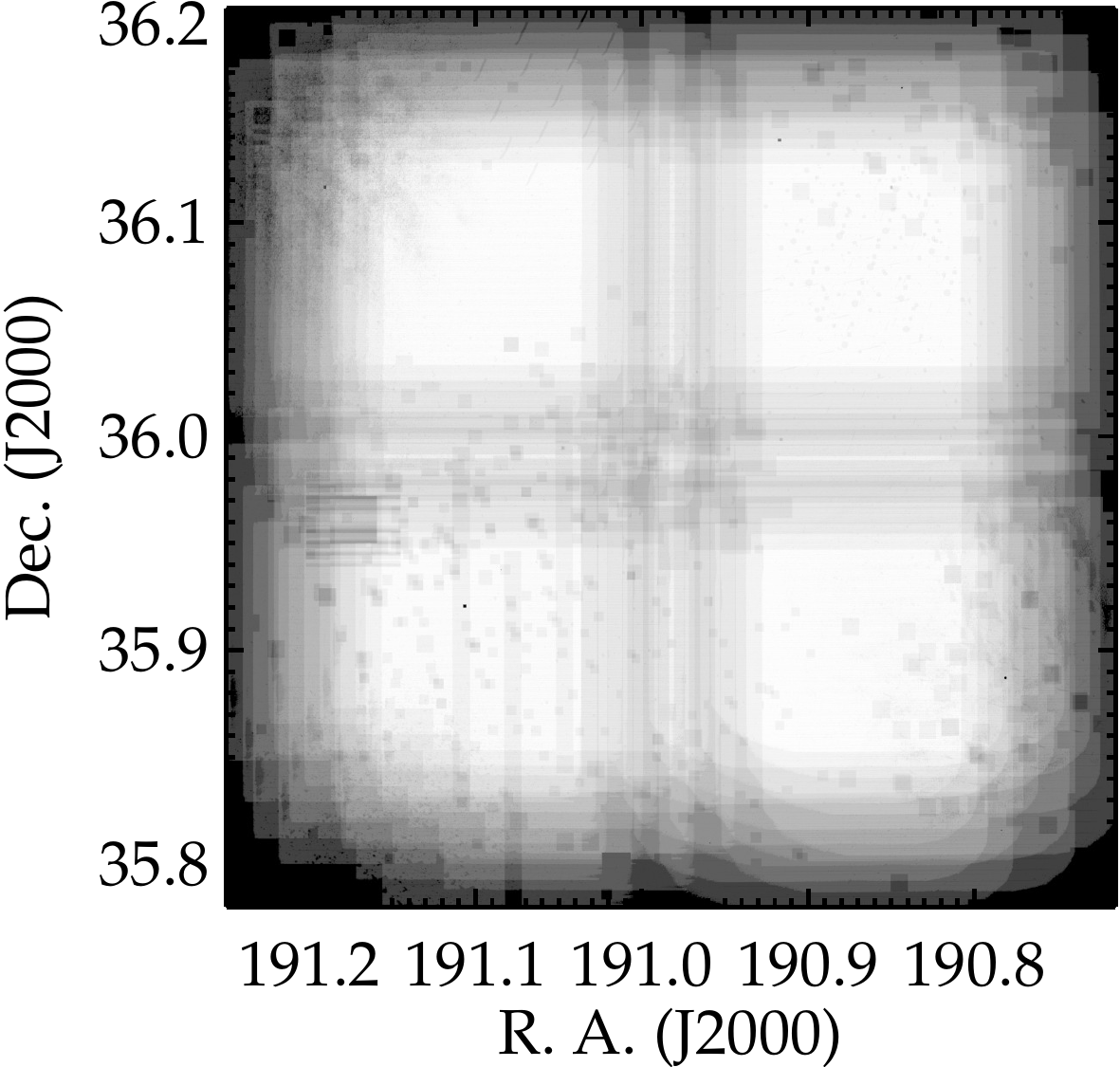}
\includegraphics[width=0.48\columnwidth]{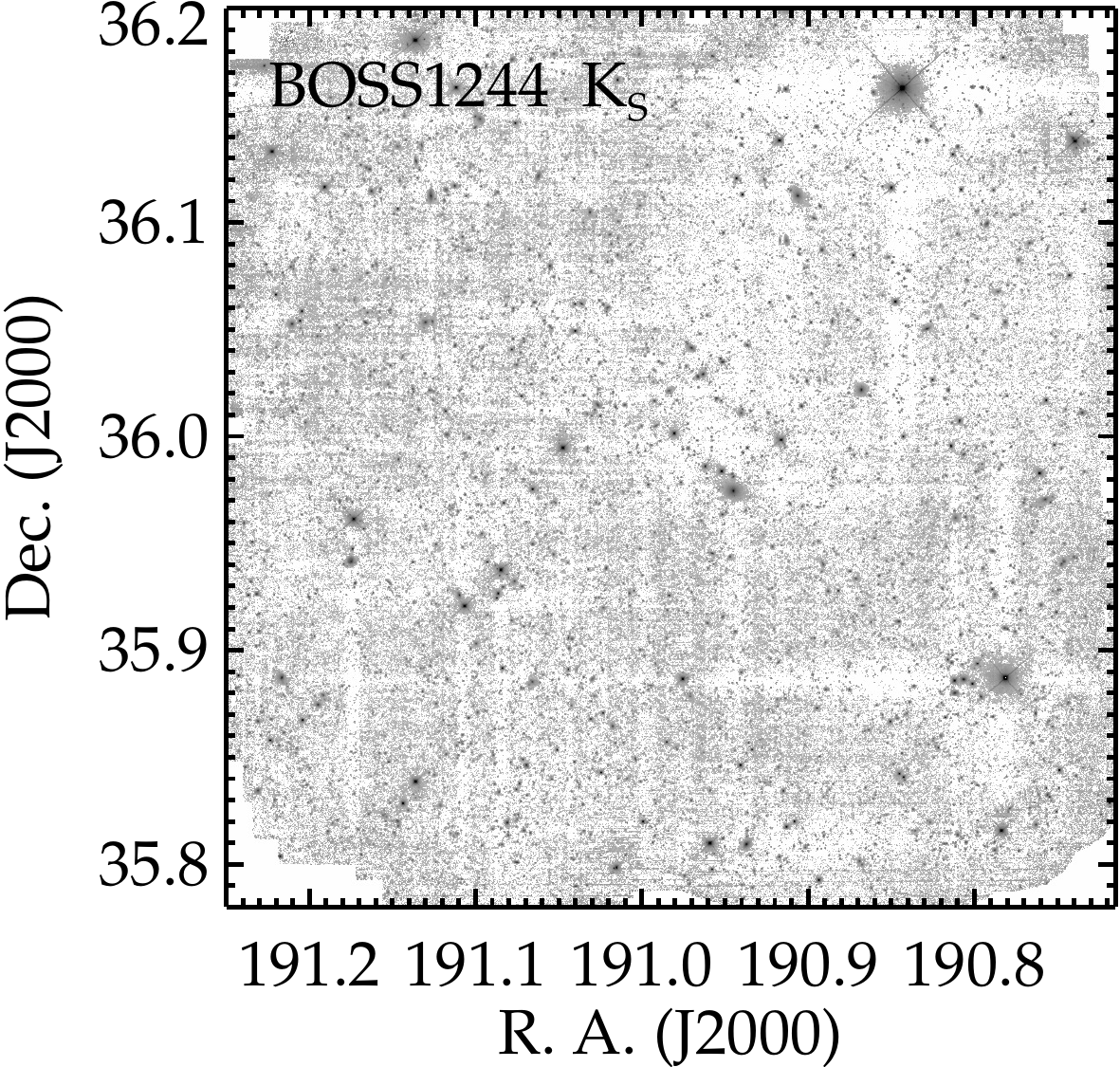}
\includegraphics[width=0.48\columnwidth]{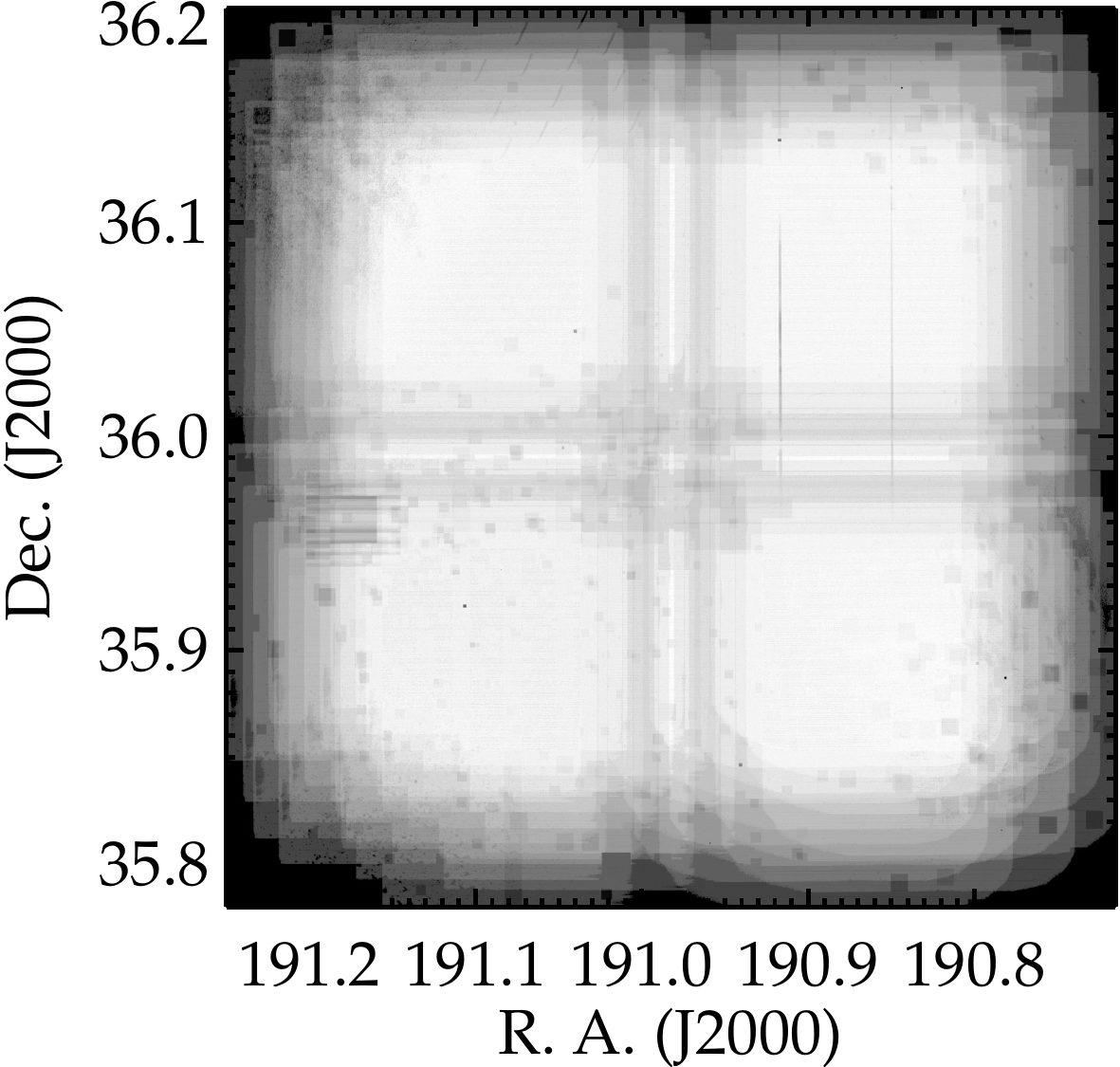}
\caption{Science image (left) and exposure map (right) taken with CFHT/WIRCam through the narrow $H_2S1$ (top) and broad $K_{\rm s}$ (bottom) filters for  the overdensity field BOSS1244.}
    \label{fig:img1244}
\end{figure}

\begin{figure}
\centering
\includegraphics[width=0.48\columnwidth]{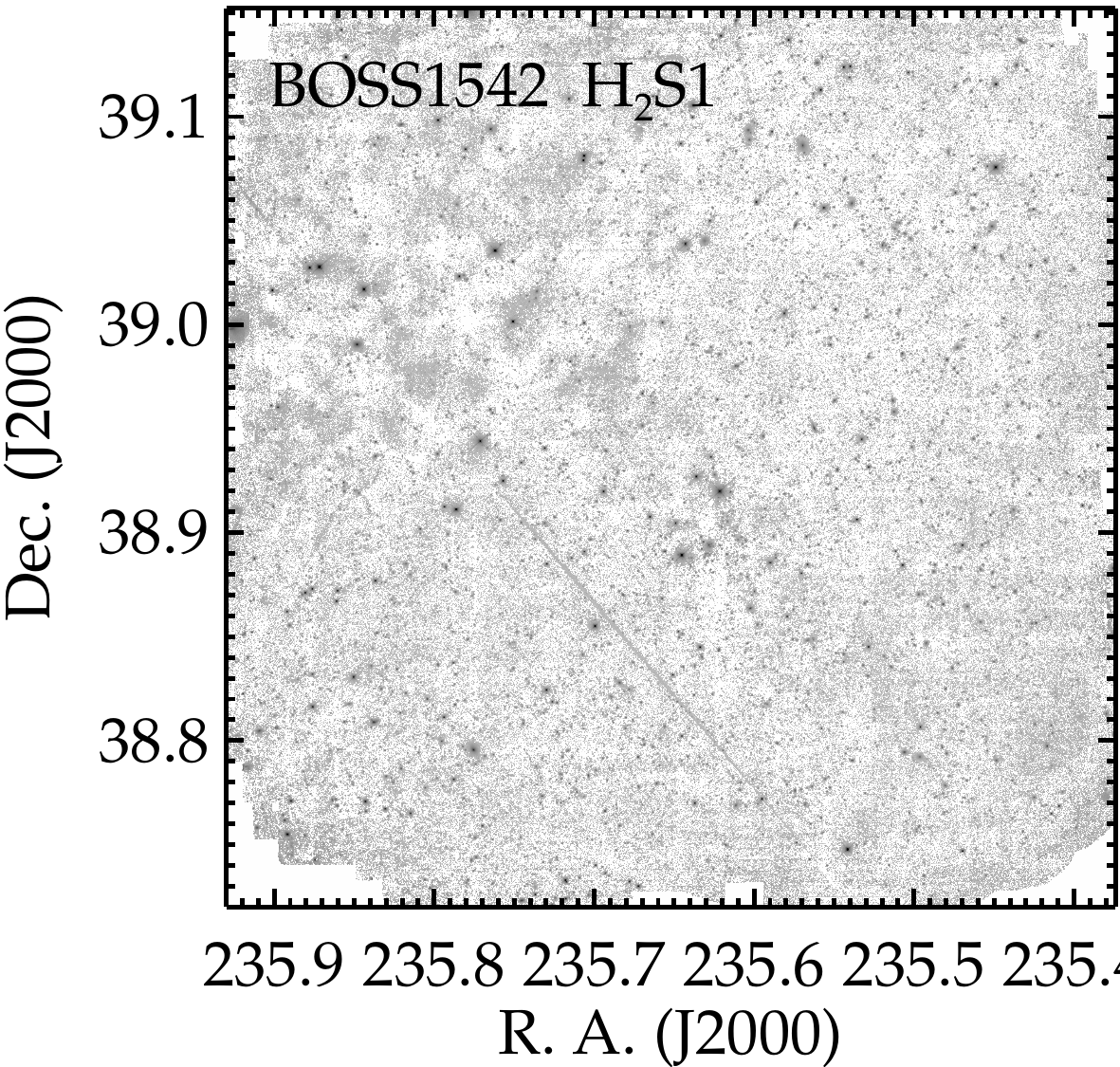}
\includegraphics[width=0.48\columnwidth]{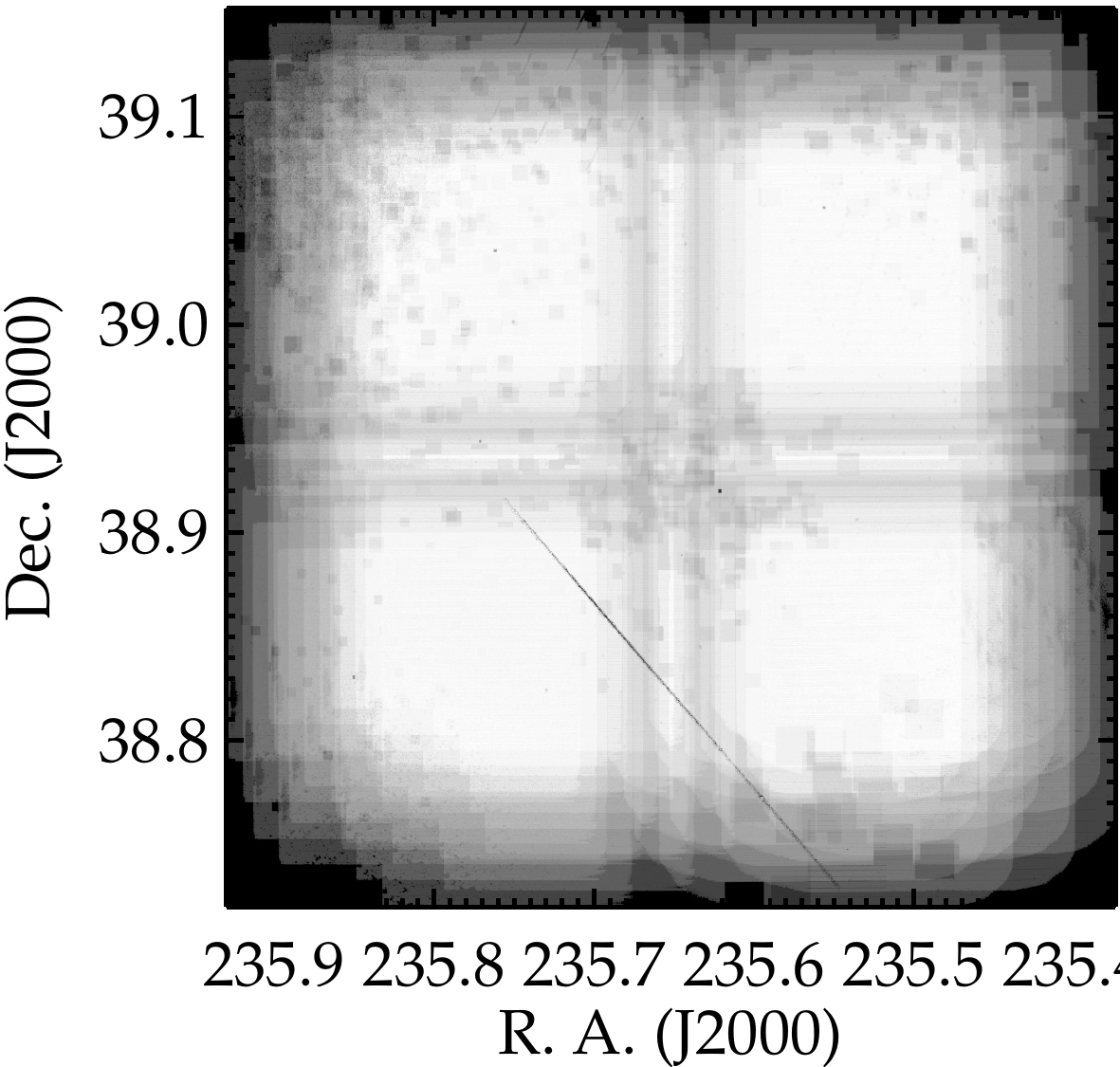}
\includegraphics[width=0.48\columnwidth]{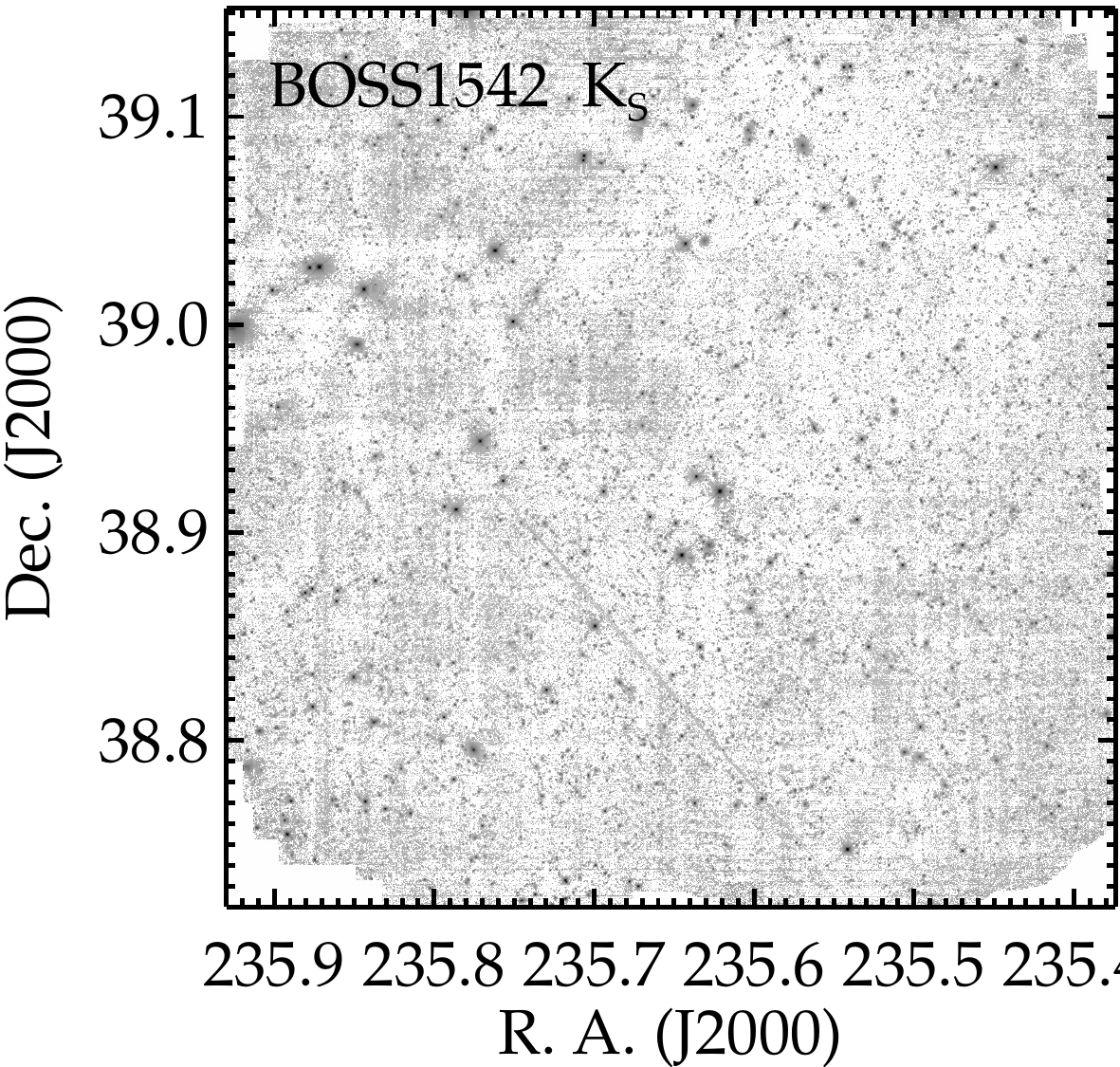}
\includegraphics[width=0.48\columnwidth]{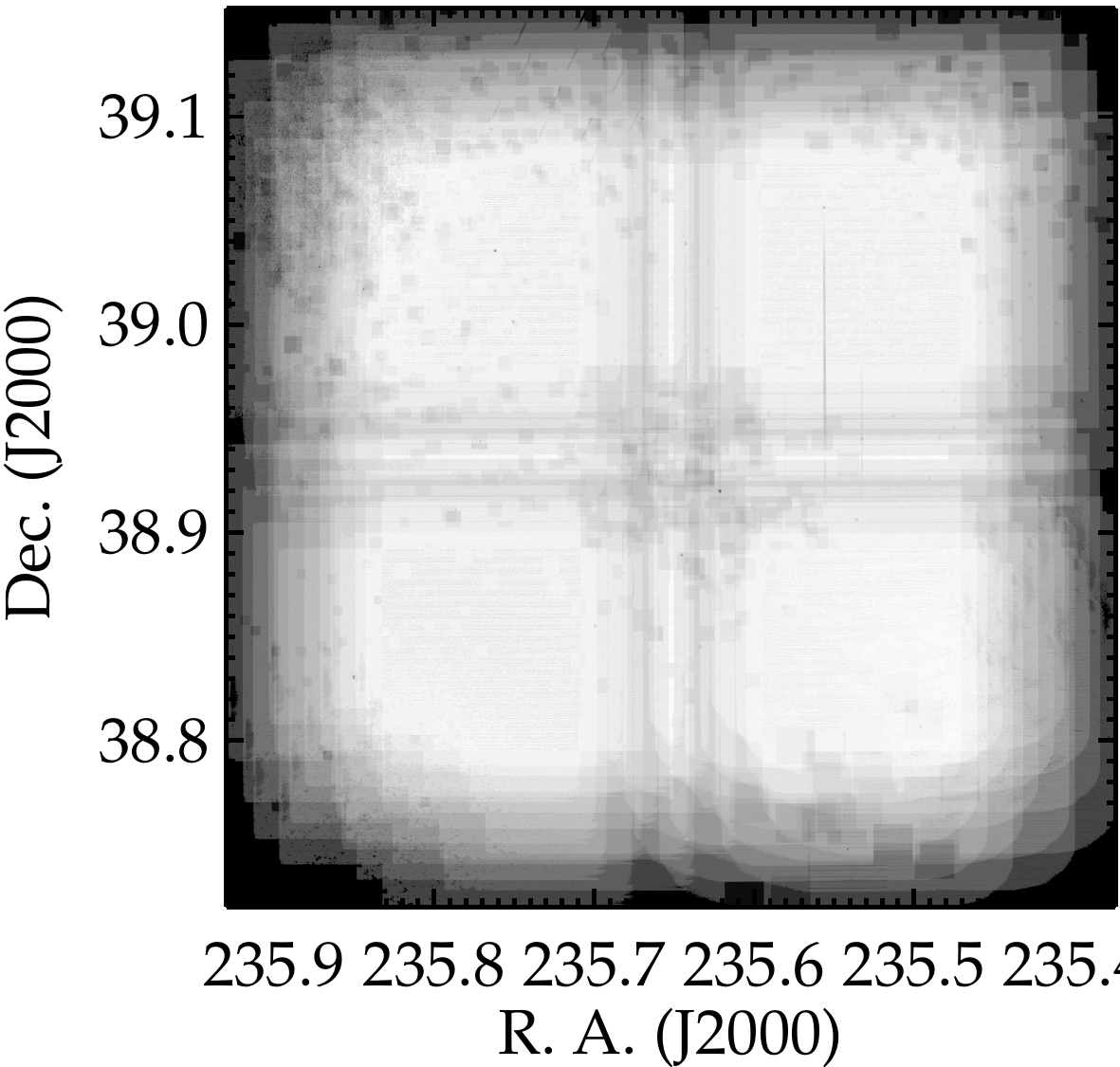}
\caption{Science image (left) and exposure map (right) taken with CFHT/WIRCam through the narrow $H_2S1$ (top) and broad $K_{\rm s}$ (bottom) filters for the overdensity field BOSS1542.}
    \label{fig:img1542}
\end{figure}

All final science images of the two MAMMOTH fields show a similar Point Spread Function (PSF) with Full Width at Half Maximum (FWHM) of 0$\farcs78\pm$0.01.  Figure~\ref{fig:img1244}  and Figure~\ref{fig:img1542} present
 the $H_2S1$ and $K_s$ science images and corresponding exposure maps for BOSS1244 and BOSS1542, respectively.  The effective area with a total integration time of $>0.5\,\times$\,maximum is 417  and 432\,arcmin$^{2}$ for $H_2S1$ and $K_{\rm s}$ in BOSS1244,   and 399 and 444\,arcmin$^{2}$ for $H_2S1$ and $K_{\rm s}$ in BOSS1542, respectively. 
The image depth (5\,$\sigma$, AB for point sources) within the effective area is estimated through random photometry on blank background using an aperture of 2$\arcsec$ diameter, giving  $H_2S1$=22.58\,mag and $K_{\rm s}$=23.29\,mag for BOSS1244 and $H_2S1=22.67$\,mag and $K_{\rm s}$=23.23\,mag for BOSS1542.

\section{Results}\label{sec:results}
 
\subsection{Identifying emission-line objects}\label{sec:iden}

 We select emission-line objects through narrow $H_2S1$ + broad $K_{\rm s}$ imaging with CFHT/WIRCam  in two 20$\arcmin\times 20\arcmin$ fields of MAMMOTH overdensities.  The software SExtractor \citep{Bertin1996} is used for source detection and flux measurement in the $H_2S1$ image.  A secure source detection is based on at least five contiguous pixels that contain fluxes above three times the background noise ($\geq 3\sigma$).  The exposure map is used as the weight image to suppress  false sources in the low signal-to-noise (S/N) area. The $H_2S1$ and $K_{\rm s}$ images are aligned into the same frame. Photometry is carried out using SExtractor under the ``dual-image'' mode, in which the flux of a source in the $K_{\rm s}$ image is measured over the same area  as in the $H_2S1$ image.  We limit source detection  in the area with a 5\,$\sigma$ depth down to $H_2S1$=22.58\,mag  for BOSS1244 and $H_2S1$=22.67\,mag for BOSS1542. The same detection area in $K_{\rm s}$ reaches a depth of $K_{\rm s}$=23.29\,mag and 23.23\,mag, respectively. In total, 6,253 and 8,012 sources are securely detected with an S/N ratio of $>5$ in the $H_2S1$ image of BOSS1244 and BOSS1542, respectively.

\begin{figure*}\centering
	\includegraphics[width=0.48\textwidth]{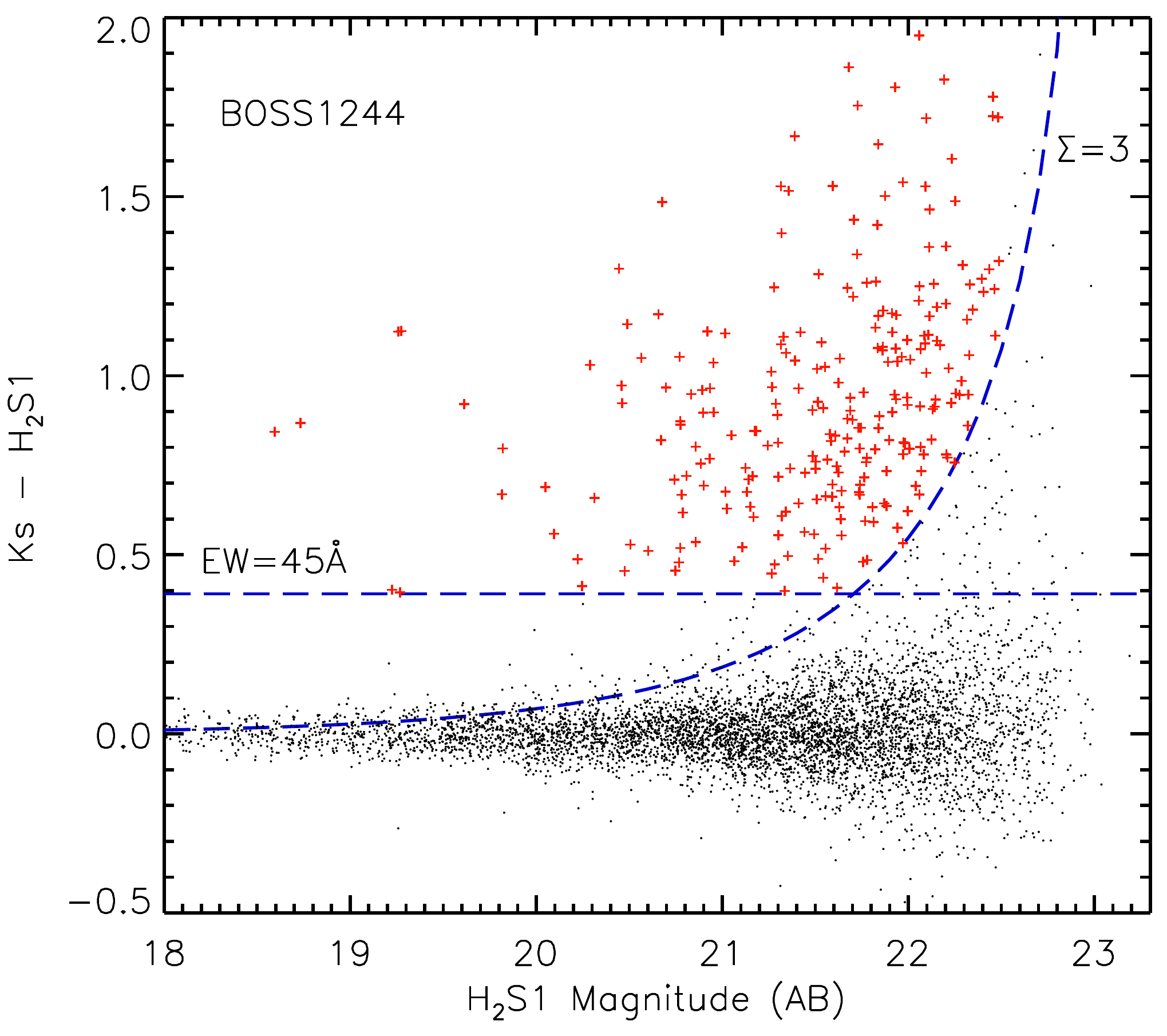}
	\includegraphics[width=0.48\textwidth]{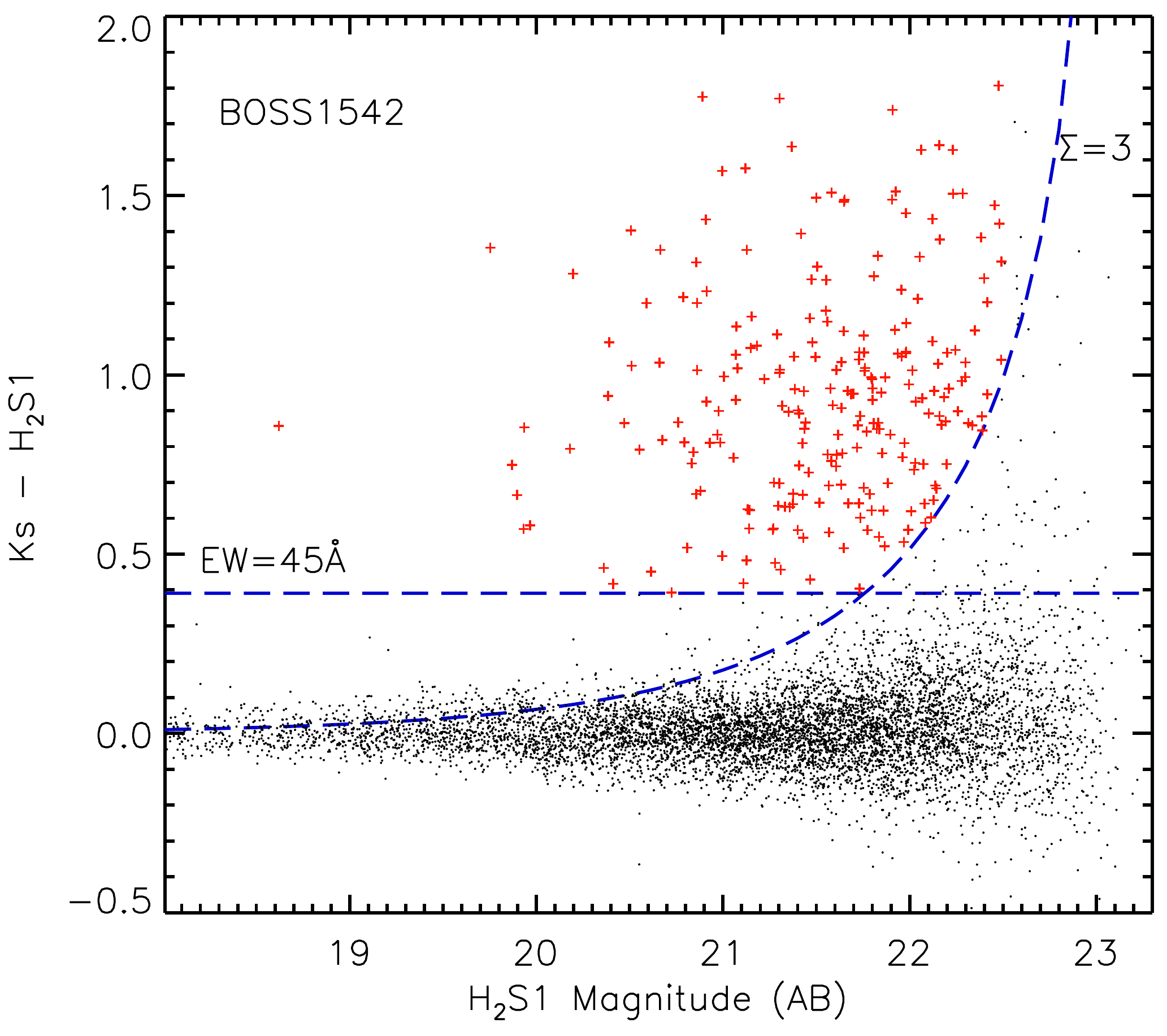}
        \caption{Selection diagram of $H_2S1$ magnitude versus $K_{\rm s}-H_2S1$  for emission-line candidates in BOSS1244 (left) and BOSS1542 (right).  The dashed curves refer to the selection limitation given in Equation~\ref{equ:sele} with a signification level of $\Sigma=3$. The horizontal dashed lines represent rest-frame $EW$=45\,\AA\ ($H_2S1 - K_{\rm s}=0.39$\,mag). The selection criteria pick up 244 (left) and 223 (right) emission-line objects with $H_2S1<22.5$\,mag.  Of them,  $\sim$196/175 are estimated to be true HAEs at  $z=2.24$ after correcting for fore- and background emitters.}
    \label{fig:selection}
\end{figure*}

The presence of a strong emission line induces a flux excess in the narrow band relative to the broad band. We use $K_{\rm s} - H_2S1$ to select emission-line objects as 
\begin{equation}  \label{equ:sele}
K_{\rm s} - H_2S1 > -2.5\,\log ( 1 - \Sigma\,\sqrt{\sigma_{K_{\rm s}}^2 + \sigma_{\rm H_2S1}^2}/f_{\rm H_2S1} ), 
\end{equation}
where $\Sigma$  is the significant factor,  $\sigma_{\rm H_2S1}$ and $\sigma_{K_{\rm s}}$ are  $H_2S1$ and $K_{\rm s}$ background noises. Here $H_2S1$-band flux  is defined as $f_{\rm H_2S1}=0.3631\times10^{0.4\,(25 - H_2S1)}$.  The background noises and $f_{\rm H_2S1}$ are given in units of $\mu$Jy.  Figure~\ref{fig:selection} shows the color $K_{\rm s} - H_2S1$ as a function of $H_2S1$ magnitude for sources detected in BOSS1244 and BOSS1542.  We adopt $\Sigma>3$ to identify emission-line objects. The strength of an emission line is quantified by the rest-frame equivalent width (EW). Here a cut of $EW>45$\,\AA\ is adopted to minimize false excess caused by the photon noises of bright objects. This cut corresponds to $K_{\rm s} - H_2S1>0.39$\,mag.  A lower EW cut (e.g., $EW>30$\,\AA) will increase only a few more candidates and thus have marginal effect on our results.

From Figure~\ref{fig:selection}, 251 and 230 emission-line candidates are selected with $\Sigma>3$, $EW>45$\,\AA\ and $H_2S1<22.5$\,mag in BOSS1244 and BOSS1542, respectively.   We visually examined these candidates and removed 7/7 of the 251/230 false sources in the two fields.  They are either spikes of bright stars or contaminations. In the end, 244 and 223 emission-line objects are identified in BOSS1244 and BOSS1542, respectively.  Among these emission-line objects, five in BOSS1244 and three in BOSS1542 are spectroscopically confirmed as QSOs at $z\sim 2.24$ in SDSS.

The emission lines in the $H_2S1$ filter may be  \Ha\ at  $z=2.24$, Pa$\alpha$ at  $z=0.14$,  \FeII\ at  $z=0.30$, Pa$\beta$ at  $z=0.66$, \SIII\ at $z =1.23/1.35$ and \OIII\ at  $z=3.25$.  By limiting the $H_2S1$ and $K_{\rm s}$ data from \citet{An2014} to the depths of our observations, we estimate that about 78 emitters would be detected over 383\,arcmin$^2$ of the Extended Chandra Deep Field South (ECDFS).  Of these emitters, $36-40$\,per\,cent  are  HAEs \citep{Hayes2010,Lee2012, An2014}, suggesting a number density of $31-34$ over 417\,arcmin$^2$ for HAEs in the general field.  The numbers of emitters we detect in the two MAMMOTH fields are much higher, undoubtably contributed by an excess of HAEs at  $z=2.24$.   This is strongly supported by the fact that a group of  CoSLAs at $z\sim 2.24$, as a convincing tracer of overdensities,  are probed by the background quasars, as well as the fact that these spectroscopically-identified QSOs are also detected as the emitters at the same redshift (i.e., $z=2.24$). Moreover, the possibility that the excess of emitters is associated with other redshift slices is negligible. We point out that the volume is too small to contain a significant number of  Pa$\alpha$ emitters at  $z=0.14$ or \FeII\ emitters at $z=0.30$. The strong \SIII\ emission lines are usually powered by shock waves in the post-starburst phase \citep{An2013}. As we will show later, the excess is contributed by an overdensity of $\delta_{\rm gal}>5$, where $\delta_{\rm gal}=(\Sigma-\Sigma_{\rm field})/\Sigma_{\rm field}$.  It is hard to believe that a large number of SFGs in such massive overdensities at  $z =1.23/1.35$ could turn them into the post-starburst phase in a locked step. The excess of emitters is unlikely associated with overdensities at $z=3.25$ traced by \OIII\ emitters because no $z\sim 3.25$ CoSLAs are found from the spectra of the background quasars. We caution that the weak overdensities at $z =1.23/1.35$ or $z=3.25$, if exist, might still contaminate the  identification of substructures in the HAE-traced overdensities at $z=2.24$ unless these emitters are identified with spectroscopic redshifts. 

\begin{figure*}
\centering
\includegraphics[width=0.48\textwidth]{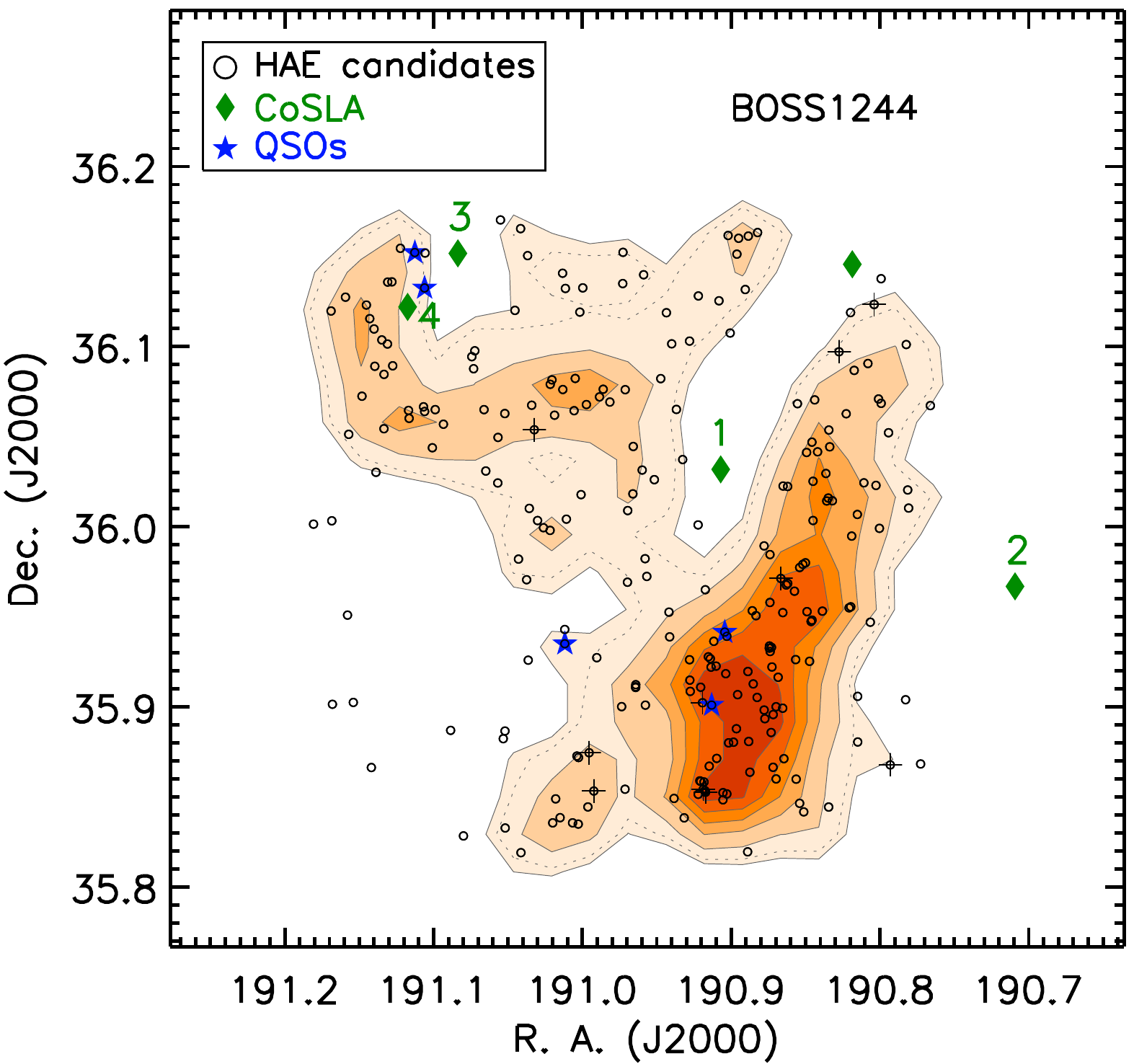}
\includegraphics[width=0.48\textwidth]{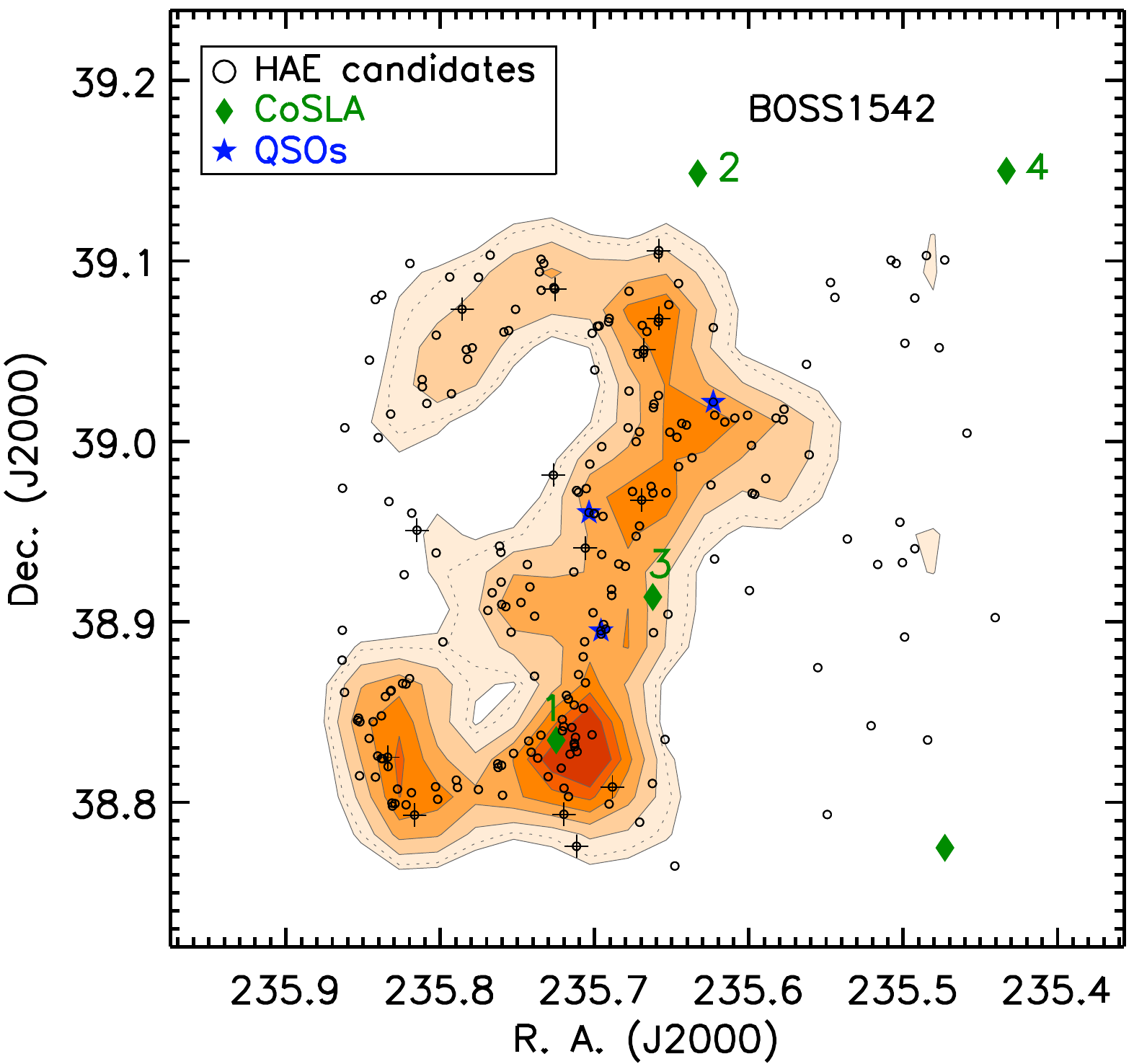}
\caption{Density maps of HAE candidates in BOSS1244 (left) and BOSS1542 (right).  Black circles represent the selected emission-line candidates.  A Gaussian kernel of $\sigma=1\arcmin$ (1.6\,cMpc at $z=2.246$) is adopted to smooth the density maps and draw the contours in the same linear scale. The contour levels refer to [4, 8, 12, 16, 20, 24]\,$\times$ the number density of HAEs in the general field (0.071 per arcmin$^2$).  The dotted lines mark the contour level of 5.2$\times$0.071 per arcmin$^2$ used as the boundaries between the outskirts and dense regions. A group of Coherently Strong Ly$\alpha$ Absorption (CoSLA, green diamonds) and QSOs (blue stars) at $z\simeq 2.24$ are marked. Green numbers mark CoSLAs shown in Figure~\ref{fig:abs1} and \ref{fig:abs2}. Black crosses pinpoint the emitters with highest line fluxes ($\log L_{\rm H\alpha}>43.4$). The HAE-traced density maps uncover that the two massive overdensities have different structures: BOSS1244 is dominated by an elongated high-density structure, and  BOSS1542 appears to be a large-scale filamentary structure.}
    \label{fig:density}
\end{figure*}

We aim to estimate the total number of HAEs detected in our fields.  It is clear that the detection rate of HAEs is sensitive to the image depths and cosmic variance. The datasets in ECDFS suggest 78  emitters (33 HAEs and 45 non-HAEs) to be detected over 383\,arcmin$^2$ using our selection criteria. We remind that this likely overestimates the emitter detection rate because the detection completeness is higher in the deeper ECDFS observations. Instead, we adopt 78 emitters detected over the survey area of 417\,arcmin$^2$ in BOSS1244, giving a detection rate of 0.187 per\,arcmin$^2$.  We adopt $36-40$\,per\,cent of the emitters as HAEs \citep{Hayes2010,Lee2012, An2014}, giving a detection rate of 0.071$\pm$0.004 per\,arcmin$^2$ for HAEs at $z=2.24$ and  0.116$\pm$0.004 per\,arcmin$^2$ for non-HAEs. We obtain 48$\pm$2 non-HAEs and 30$\pm$2 HAEs over the same area in the general field. We use these two numbers for both of our two fields and ignore the variation in survey area.  Of 244/223 emission-line objects, we estimate the number of HAEs at $z=2.24$ to be $196\pm 2$/$175\pm 2$ in BOSS1244/BOSS1542,  yielding an overdensity factor $\delta_{\rm gal}=5.6\pm 0.3$ for BOSS1244 and $4.9\pm 0.3$ for BOSS1542.  Here the errors account only for the variation in the fraction of HAEs in the general field. The uncertainty in the detection rate of emitters is mostly driven by the cosmic variance  and not counted here.  We notice that there are only 21/28 objects in the low-density regions of  BOSS1244/BOSS1542, giving an emitter detection rate of 0.124/0.135 per\,arcmin$^2$ (see next section for more details) slightly lower than the adopted value (0.187 per\,arcmin$^2$). This hints that the cosmic variance may induce an uncertainty up to 50\,per\,cent. Since the fraction of HAEs in the low-density regions is unknown, we choose ECDFS as a representative for the general field. We point out that decreasing the detection rate for the general field will increase the overdensity factors that we estimated, and further strengthen our conclusions.  When focusing on the high-density regions (see Figure~\ref{fig:density}), the overdensity factor increases by $2-3$ times.

The redshift slice of  $z=2.246\pm 0.021$ over $20\times 20$\,arcmin$^2$ corresponds to a co-moving box of $54.3\times 32.0 \times 32.0$ (=55,603)\,$h^{-3}$\,cMpc$^3$, equal to a cube of 38.2\,$h^{-1}$\,cMpc each side.  The overdensity factor of $\delta_{\rm gal}\sim 5-6$ over such a large scale displays the overdensities in our two target fields as most massive ones at the epoch of $z\sim 2-3$ \citep{Cai2016}.  We thus conclude that the large excess of HAEs confirms BOSS1244 and BOSS1542 as massive overdensities at  $z=2.24$. The confirmation validates the effectiveness of the MAMMOTH technique in identifying the massive overdensities of scales $15-30\,h^{-1}$\,Mpc at $z\sim 2-3$. 
We note that the redshift slice is given by the width of the $H_2S1$ filter that corresponds to a line-of-sight distance of 54.3\,cMpc at  $z=2.24$.  This scale should be sufficiently large for the detection of the progenitor of local clusters like the Coma \citep{Chiang2013}, although there is still possibility that some galaxies of the overdensities might spread out of the redshift slice  (e.g., a protocluster across $\sim 60$\,cMpc in the SSA22 field; \citealt{Matsuda2005}) possibly  partially due to the Fingers of God effect caused by the peculiar velocities of galaxies.  A complete census of these massive overdensities will require spectroscopic surveys of galaxies over a larger sky coverage and wider range in redshift to map the kinematics of the overdensities and their surrounding density fields.

\subsection{Density maps of H$\alpha$ emitters}\label{sec:density}

 We estimate that 196 of 244 (80\,per\,cent) and 175 of 223 (78\,per\,cent) emission-line objects are HAEs at  $z=2.24$ that belong to the massive overdensity BOSS1244 and BOSS1542, respectively.  The non-HAEs are located at fore- or background of the  $z=2.24$ slice.  In the ECDFS field, non-HAEs consist of 42\,per\,cent foreground and 58\,per\,cent background emitters (i.e., \OIII\ emitters at $z=3.25$), when limiting the detection to the depths of the BOSS1244 observations. One can expect that these non-HAEs spread randomly over the observed area.  We thus use all emitters to build the density map and the presence of non-HAEs can be seen as a flat density layer in a statistic manner.  We adopt the number density of 0.116 per arcmin$^2$ for non-HAEs and  of 0.071 per arcmin$^2$ for  $z=2.24$ HAEs in the general field.  We note, however,  that galaxies reside in the cosmic web, we can not exclude the possibility that the foreground or background non-HAEs might associate with some structures and  contaminate the density maps of HAEs at $z=2.24$.

We identified the emission-line objects in the high-$S/N$ regions, corresponding to an area of $20\arcmin \times 20\arcmin$ in each field.  We treat each object equally and use the  projected number density of all emitters to trace the projected matter density. The detection area is divided into a grid of cells with $1\farcm2\times 1\farcm2$ each, and the number of emitters in each cell region is then counted to generate a density map.  A Gaussian kernel of $\sigma$=1$\arcmin$ (1.6\,cMpc at $z=2.246$) is utilized to convolve the density map. Contours of density maps are drawn at the levels of 4, 8, 12, 16, 20 and 24\,$\times$ the number density of HAEs of the general field  (0.071 per arcmin$^2$) plus the number density of non-HAEs (0.116 per arcmin$^2$). Figure~\ref{fig:density} shows the spatial distributions of emission-line objects in two MAMMOTH fields, over plotted with the density maps.  The contours in two density maps are given in identical levels in order to compare these two fields. Out of the first contour level lines, there are only 21/28 objects over 169/207\,arcmin$^2$ in BOSS1244/BOSS1542, giving an emitter detection rate of 0.124/0.135 per arcmin$^2$ in 41/52\,per\,cent of the total area of the two fields. These indicate that the number density of non-HAEs ($\sim$60\,per\,cent of the total) is likely significantly lower than that in ECDFS.

It is clear from Figure~\ref{fig:density} that the density maps traced by HAEs reveals sub-structures of the two massive overdensities. BOSS1244 exhibits two components within the observed area --- a low-density component connected to an elongated high-density component of a scale of 25$\times$10\,cMpc.  The high-density component spreads over an area of 103\,arcmin$^2$ within the first contour level and reaches an overdensity factor of $\delta_{\rm gal}\sim$15, and even $\sim$24 in the central 4$\arcmin\times 6\arcmin$ region.  The massive overdensities of $\delta_{\rm gal}>6$  traced by star-forming galaxies over (15\,cMpc)$^3$ are predicted by simulations exclusively to be proto-clusters, i.e., the progenitors of massive galaxy clusters of $>10^{15}$\,M$_\odot$ (e.g., Coma cluster) in the local universe \citep{Chiang2013}.  One caveat is that the elongated structure in BOSS1244 might be extended or divided into multiple components along line of sight over 54.3\,cMpc. Even if we divide the overdensity factor $\delta_{\rm gal}\sim$15 by three to match the volume of (15\,cMpc)$^3$, the divided structures would be still sufficiently massive and overdense to form massive clusters.

In contrast, BOSS1542 can be seen as a large-scale filamentary structure with multiple relatively dense clumps.  The density and size of these clumps are significantly smaller than the dominant component of BOSS1244. The total length of the structure along the filament reaches 50\,cMpc. This structure covers an area of 192\,arcmin$^2$ ($\sim 32\times15$\,cMpc at $z=2.246$) and yields $\delta_{\rm gal}\sim$10 within the first contour level shown in Figure~\ref{fig:density}.  The bottom part at $Dec.<38.88$ spread over 72\,arcmin$^2$ ($\sim 12\times15$\,cMpc at $z=2.246$) and have a mean $\delta_{\rm gal}\sim$11.  We suspect that at least part of the filamentary structure could  eventually condense into one massive galaxy cluster as revealed by simulations \citep[e.g.,][]{Chiang2013}.  
Spectroscopic observations can map kinematics of member galaxies and quantitatively determine if different components in these overdensities could merge into one mature cluster of galaxies.   We will carry out a detailed analysis of dynamics and masses for the two HAE-traced overdensites using spectroscopic data in a companion work (Shi D.~D. et al., in prep).

\begin{figure*}
\centering
\includegraphics[width=0.48\textwidth]{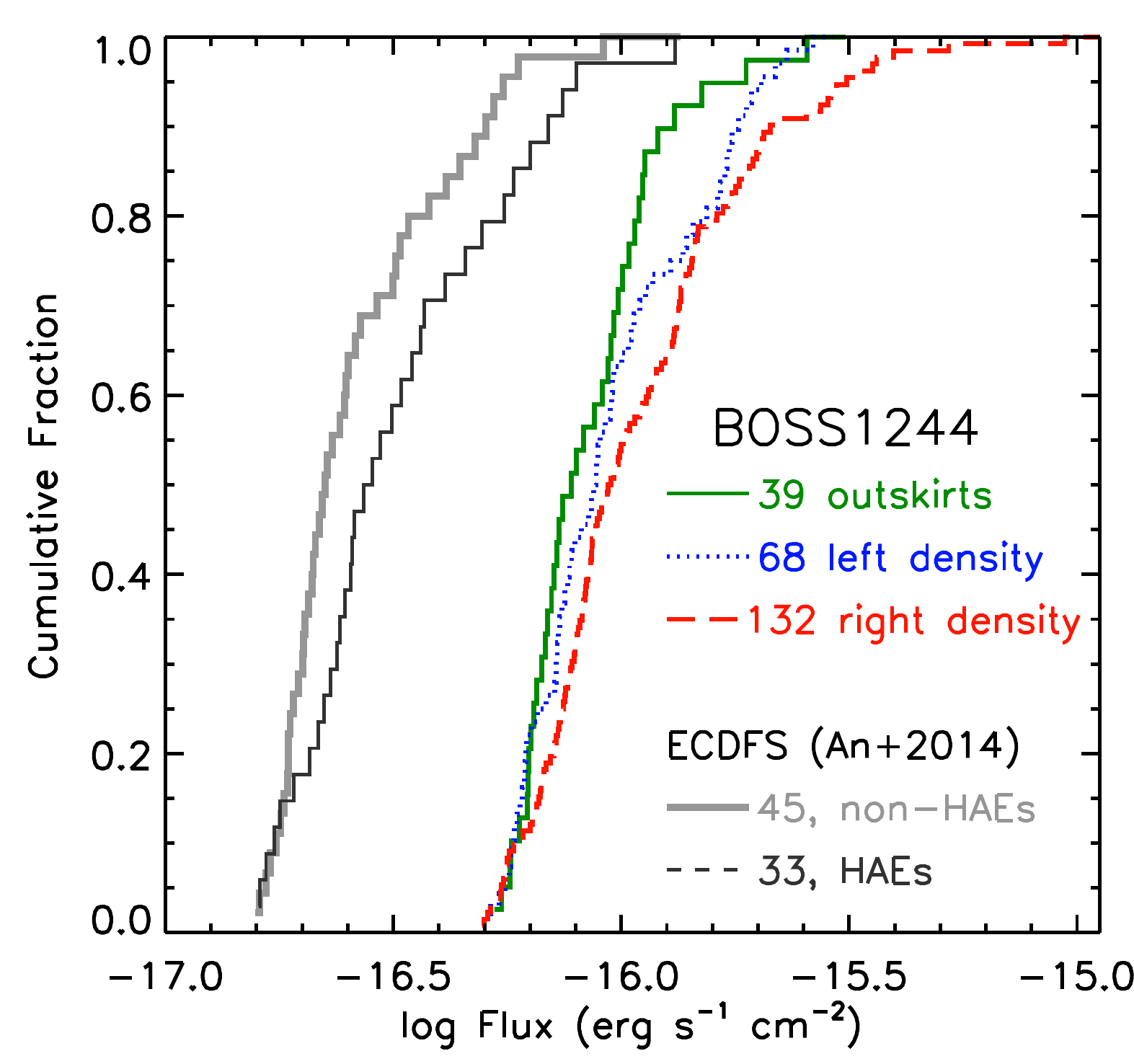}
\includegraphics[width=0.48\textwidth]{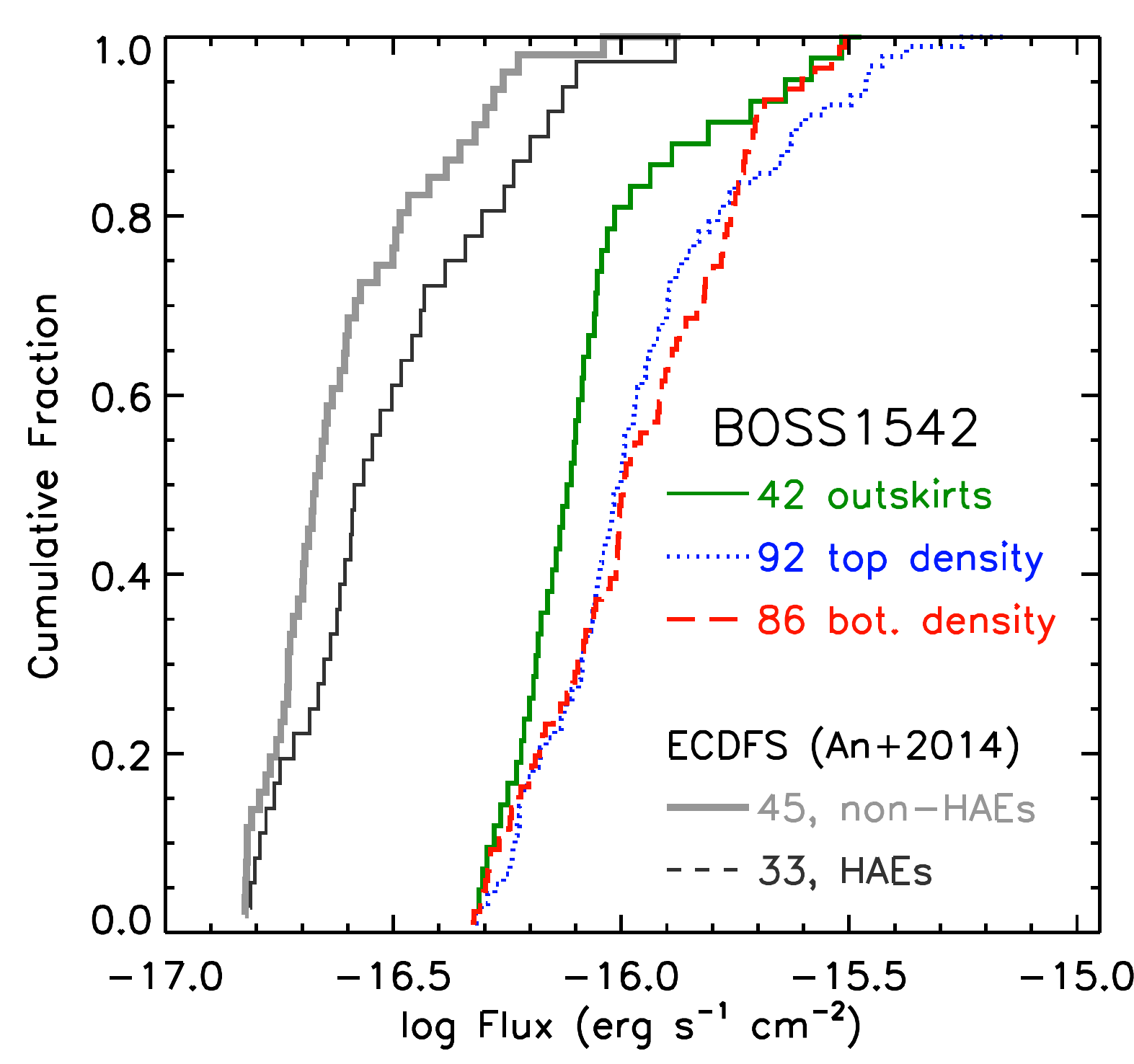}
\caption{Cumulative fraction of the observed line fluxes in different parts of BOSS1244 (left) and BOSS1542 (right).  The description of these parts are given in the text. The numbers of line emitters in these parts are presented.  The line emitters in the ECDFS field from \citet{An2014} are taken as the representative of the general field at $z=2.24$.  Here the detection in ECDFS is limited to the same depths of the $H_2S1$ and $K_{\rm s}$ observations in BOSS1244. The cumulative curves of \Ha\ emitters (black) and of non-\Ha\ emitters (gray) in ECDFS are shifted by $-$0.5\,dex for clarity. We can see that the line fluxes of \Ha\ emitters are relatively higher (by $\sim$0.1\,dex) than those of non-\Ha\ emitters in ECDFS.   In BOSS1244 and BOSS1542, the emitters in the high-density regions also have emission lines relatively brighter than the emitters in the outskirts. We note that a correction of \NII/\Ha=0.117 is needed to obtain \Ha.  The flux of \Ha\ at $z=2.24$ can be converted to luminosity by adding 58.58\,dex and  extinction correction for $A(\rm H\alpha)=1$\,mag will increase the luminosity by 0.4\,dex. } \label{fig:dist} 
\end{figure*}

We notice that the density maps of our two HAE-traced structures might be contaminated by the fore- or background emitters that are probably associated with overdensities.  We examine the possibility by comparing the line flux distributions of those emitters in dense regions and outskirts.  Out of the first contour level lines, there are only 21/28 objects in BOSS1244/BOSS1542. We adopt the contour lines at the level of 5.2$\times$0.071 per arcmin$^2$ (the dotted lines in Figure~\ref{fig:density})  as the boundaries of the dense regions to ensure that the outskirts contain $\sim$40 objects sufficient  for a meaningful statistics and avoid serious contamination from the dense regions at the same time. In BOSS1244, the dense regions include two parts: left density and right density (i.e., the elongated structure). In BOSS1542, HAEs form a giant filamentary structure. We split the dense regions  into two roughly equal parts via a horizontal line at $Dec=38.92$: top density and bottom density. Figure~\ref{fig:dist} shows the cumulative curves of line fluxes in different parts of the two MAMMOTH fields.  Here five QSOs in BOSS1244 and three QSOs in BOSS1542 are excluded. For comparison, we present the cumulative curves for the HAEs at $z=2.24$ and non-HAEs (i.e., fore- and background emitters) in ECDFS from \citet{An2014}. These emitters are selected at the same detection depths as our BOSS1244 observations.  

It is clear from Figure~\ref{fig:dist} that in ECDFS the observed line fluxes of the HAEs at $z=2.24$  are systematically higher by 0.1\,dex than those of the non-HAEs. We note that the 45 non-HAEs include 26 \OIII\ emitters at $z=3.25$ that appear globally fainter than HAEs at $z=2.24$. In BOSS1244, the right density (i.e., the elongated dominant structure in Figure~\ref{fig:density}) contains emitters with line fluxes relatively higher than the emitters in the outskirts; the left density shows a cumulative curve similar to that of the outskirts. It is worth noting that the left density made of 68 objects are an extended and weak concentration, and the right density host all emitters with $\log f_{\rm line} > -15.6$.   In BOSS1542, the line fluxes of the emitters in both the top and bottom density are statistically higher by typically $\sim$0.1\,dex in comparison with those of the emitters in the outskirts; the top density contains more objects with high line fluxes.  We can conclude that the emitters in the high-density regions have line fluxes globally higher than the emitters in the outskirts of the two MAMMOTH overdensities, following the difference of the line flux distributions between HAEs and non-HAEs in ECDFS.

Moreover, the cumulative distribution of line fluxes of the emitters in the outskirts appears analogous to that of the non-HAEs in ECDFS. We use Kolmogorov$-$Smirnov (K$-$S) test to quantify the probability that two samples are drawn from the same population. It measures the significance level of consistency of two cumulative distributions.  The $p$-value of K$-$S test is 0.59, 0.06 and 0.001 when comparing the non-HAEs in ECDFS with the emitters in the outskirts, left-density and right-density regions of BOSS1244, and 0.38, 1.45$\times10^{-5}$ and 5.40$\times10^{-5}$ with these in the outskirts, top-density and bottom-density regions of BOSS1542, respectively.  Similarly, K$-$S test yields 0.26, 0.96 and 0.97 for the HAEs in ECDFS in comparison with the three emitter samples in BOSS1244, and 0.19, 0.22 and 0.24  with the three emitter samples in BOSS1542, respectively.  
These results show that the emitters in the outskirts of two MAMMOTH overdensities satisfy the line flux distribution of the non-HAEs in ECDFS at a high significance level, and their line flux distribution inevitably differs from that of the HAEs.  On the other hand, the emitters in the high-density regions exhibit similar line flux distribution to the HAEs in ECDFS. The consistency is weaker for BOSS1542 because this giant filamentary structure contains more emitters with high line fluxes. These results support our conclusion that the high-density structures are dominated by \Ha\ emitters at $z=2.24$ and unlikely significantly contaminated by fore- or background emitters.  Again,  spectroscopic observations will play a key role in characterizing  these density substructures. 

The large-scale overdensities found at $z>2$ often exhibit filamentary structures or multiple components. The   $z=3.1$ overdensity in the SSA22 field consists of three extended filamentary structures \citep{Matsuda2005,Yamada2012}; \citet{Lee2014} reported a structure over 50\,cMpc containing multiple protoclusters at  $z=3.78$ in the Bo\"{o}tes field and these protoclusters are connected with filamentary structures;  a multi-component proto-supercluster at  $z=2.45$ has been found in the COSMOS field,  expanding over $> 60$\,cMpc in all three dimensions \citep{Cucciati2018}; the massive protocluster at  $z=3.13$ in the D1 field of the CFHT Legacy Survey (CFHTLS) also exhibits multiple density components traced by LAEs and LBGs \citep{Toshikawa2016,Shi2019}. 
The first overdensity discovered using the MAMMOTH technique, BOSS1441at  $z=2.32$, is an elongated large-scale structure of LAEs on a scale of 15\,cMpc \citep{Cai2017}. Traced mostly by LAEs or LBGs, these large-scale structures represent the extremely massive overdensities at $z \sim 2-4$.  In simulations large-scale overdensities of multiple components at $z>2$ are found to be very rare,  being solely the progenitors of massive structures of $\sim$10$^{15}$\,M$_\odot$ \citep{Topping2018}.

\begin{figure*}
\centering
\includegraphics[width=0.48\textwidth]{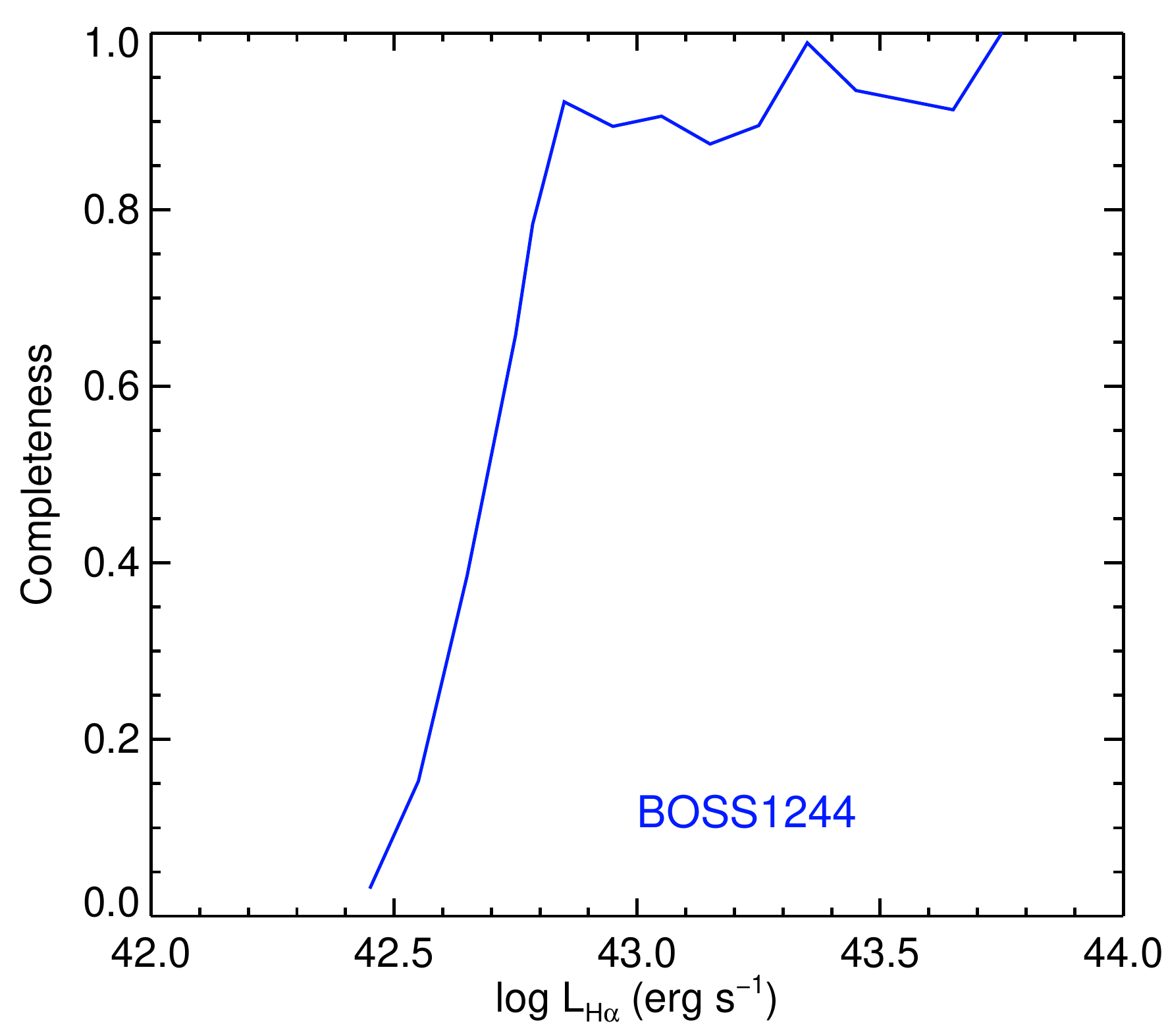}
\includegraphics[width=0.48\textwidth]{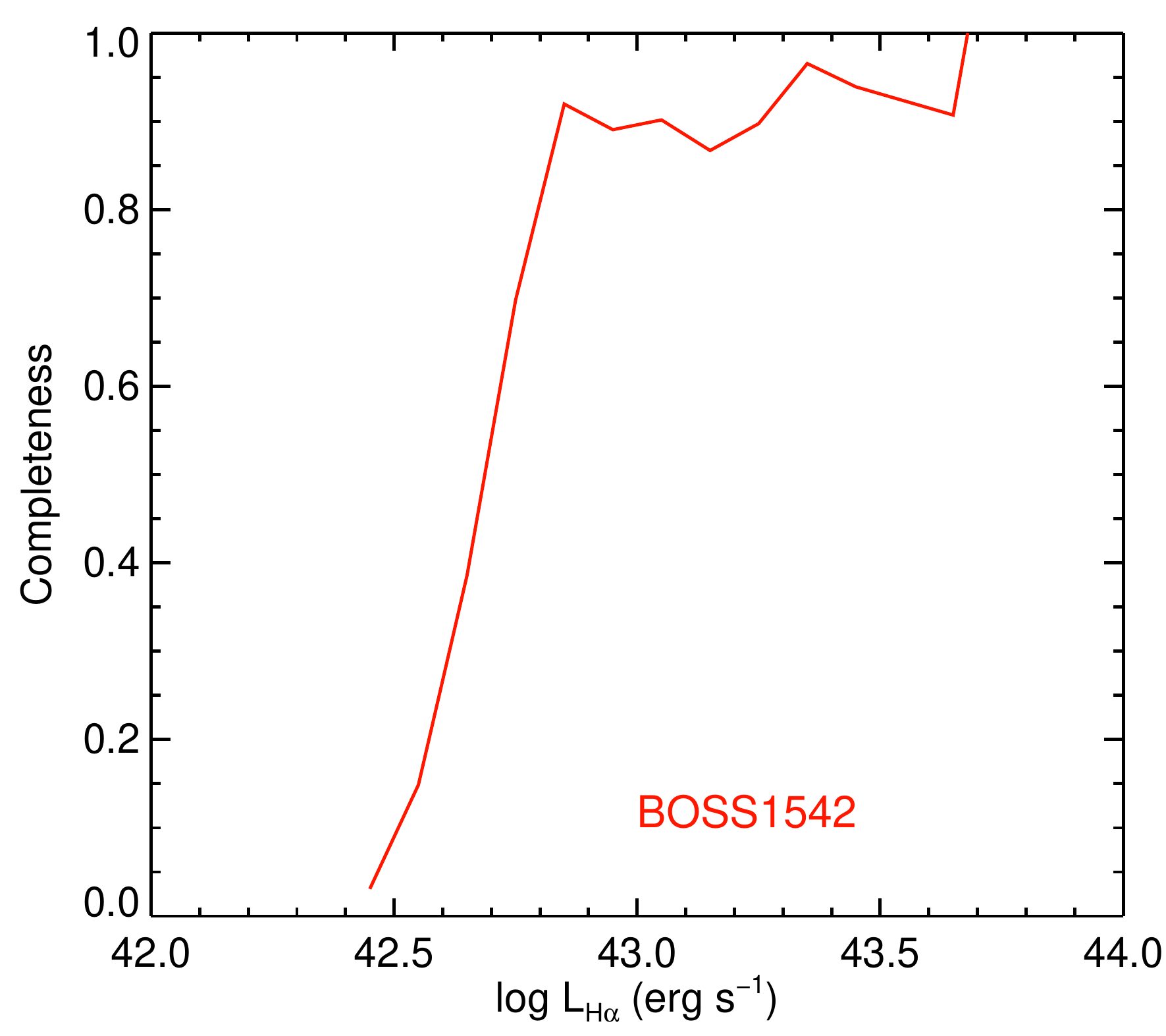}
\caption{Completeness as a function of the intrinsic \Ha\ luminosity in BOSS1244 (left) and in BOSS1542 (right). } \label{fig:lfcom}
\end{figure*}

\subsection{H$\alpha$ luminosity function}\label{sec:function}

It is essential to derive the luminosity function (LF) of the intrinsic \Ha\ luminosity that can be used as an SFR indicator of a galaxy.  This will allow us to make a direct comparison of our overdensities with the general field and examine the distribution of star formation in member galaxies of the overdensities. Below we describe the procedure for building the \Ha\ LF. This is done in the same way for both BOSS1244 and  BOSS1542. 

\subsubsection{Estimate of H$\alpha$ luminosities}

We calculate \Ha+\NII\ flux density (erg\,s$^{-1}$\,cm$^{-2}$) from the narrowband excess $K_{\rm s} - H_2S1$ and $K_{\rm s}$ total magnitude using the formula 
\begin{equation}  
 F=\Delta H_2S1 \times \frac{f_{H_2S1} - f_{Ks}}{1 - \Delta H_2S1/\Delta K_{\rm s}},
\end{equation}
where $f_{H_2S1}$ and $f_{Ks}$ refer to flux densities given in the units of erg\,s$^{-1}$\,cm$^{-2}$\,\AA$^{-1}$ in the $H_2S1$ and $K_{\rm s}$ bands  with band widths $\Delta H_2S1$=293\,\AA\ and $\Delta Ks$=3250\,\AA, respectively. Following \citet{An2014}, we use \NII/\Ha=0.117 to subtract the contribution of \NII$\lambda\lambda$6548, 6583 and obtain the observed \Ha\ line flux. 
The selection cut $EW > 45$\,\AA\ (i.e., $K_{\rm s}-H_2S1 > 0.39$\,mag) together with the 5\,$\sigma$ depths of $H_2S1$=22.6\,mag and $K_{\rm s}$=23.3\,mag (BOSS1244) determines an \Ha\ flux detection limit of $> 2.5\times 10^{-17}$\,erg\,s$^{-1}$\,cm$^{-2}$. We adopt $D=17,892$\,$h^{-1}$\,Mpc as the luminosity distance  to  $z=2.246$ to convert the \Ha\ line flux into the observed \Ha\ luminosity  for all \Ha\ emitters.  
Following \citet{Sobral2013} a constant extinction correction $A$(H$\alpha$)=1\,mag is applied to obtain the intrinsic \Ha\ luminosity, which is used to construct the \Ha\ luminosity function.

We derive SFR from the intrinsic \Ha\ luminosity following  $\log(\sfr/{\rm M_\odot}$\,yr$^{-1}$)=$\log (L_{\rm H\alpha}) - 41.27$ given in \citet{Kennicutt2012}.  The \Ha\ flux detection limit corresponds to an SFR of 5.1\,M$_\odot$\,yr$^{-1}$.   

\subsubsection{The intrinsic EW distribution}

Next step is to derive the completeness across the intrinsic \Ha\ luminosity  through fully accounting for detection limits and photometric selection.  As shown in Figure~\ref{fig:selection}, our sample selection is done with the $K_{\rm s}-H_2S1$ excess (i.e., an EW) together with source magnitudes in the two bands.  We realize that HAEs of a given \Ha\ luminosity can be bright with low EWs or faint with large EWs.  We thus need to know the intrinsic EW distribution and  quantify the noise effects on our sample selection. 
A log-normal distribution of EW is adopted for the observed \Ha +\NII\ \citep{Ly2011} to conduct Monte Carlo simulations and estimate completeness for individual \Ha\ luminosity bins of our data. 

To determine the intrinsic EW distribution of \Ha +\NII\ fluxes of our sample HAEs, we use a method based on a maximum likelihood algorithm \citep[see][for more details]{An2014}. We generate log-normal EW distributions having the mean $\log (EW_{\rm rest}$) ranging between 1.8 and 2.3 and the dispersion $\sigma [\log ( EW_{\rm rest}$/\AA)] ranging between 0.15 and 0.65 with a step of 0.1\,dex for both parameters.  We assume that \Ha+\NII\ flux is uncorrelated with its EW. This allows us to produce $H_2S1$ and $K_{\rm s}$ magnitudes by randomly assigning  EWs that obey a given distribution to the observed \Ha+\NII\ fluxes. Accounting for the background noises from our $H_2S1$ and $K_{\rm s}$ images, we apply the $H_2S1-K_{\rm s}$ selection criteria to the simulated galaxies. For  each of input intrinsic EW distributions, the `observed' EW distribution is modeled to match our $H_2S1$ and $K_{\rm s}$ observations.  We determine the intrinsic EW distribution best matching the observed EW distribution of our sample HAEs from the modeled EW distributions using the least-square method.  The best-fitting EW distribution is described by a mean $\log$($EW_{\rm rest}$)=2.00  and a dispersion $\sigma$[$\log$($EW_{\rm rest}$/\AA)]=0.35.

\subsubsection{Deriving the detection completeness}

We use the Monte Carlo simulation method to generate mock catalogs of \Ha\ emission-line galaxies satisfying a given \Ha\ LF at  $z=2.24$. The mock catalogs are used to derive the detection completeness after accounting for the noises and detection limits in our $H_2S1$ and $K_{\rm s}$ observations.  We adopt the Schechter function with $L_{\rm H\alpha}^{\ast}$=10$^{42.88}$, $\alpha$=$-$1.60 and $\log\phi^{\ast}$=$-$1.79 from \citet{Sobral2013} as the intrinsic \Ha\ LF for our two overdensities.   \citet{An2014} pointed  out that the \Ha\ LF has a shallower faint-end slope ($\alpha$=$-$1.36) and mirrors the stellar mass function of SFGs at the same redshift. They derived extinction correction for individual HAEs and recovered some heavily-attenuated HAEs that appear to be less luminous from the observed \Ha\ luminosity.  However, we are currently unable to derive extinction for individual \Ha\ emitters because of the lack of multi-wavelength observations. In \citet{Sobral2013} a constant correction $A({\rm H\alpha})$=1\,mag was applied for all \Ha\ galaxies.  We adopt their \Ha\ LF and extinction correction in our analysis.

The $H_2S1$ filter centers at $\lambda_{\rm c}=2.130\,\micron$ with an effective width of $\Delta \lambda =0.0293\,\micron$, and probes \Ha\ in a redshift bin of  $2.225<z<2.267$. 
We use this redshift bin to compute the effective volume for our sample.  The extended wing of the filter transmission curve may allow brighter emission-line objects to be detectable than the faint ones. We thus simulate HAEs over $2.20<z<2.29$ in order to estimate the contribution of the HAEs  out of  the redshift bin $2.225<z<2.267$.  The redshift span of $2.20<z<2.29$ is divided into 30 bins. In each redshift bin, one million mock galaxies are generated to have \Ha\ luminosities spreading into 500 bins  between $40 < \log (L_{\rm H\alpha}$/erg\,s$^{-1}) < 50$ and following the given LF. There are typically $\sim$2000 simulated galaxies in each bin. A flux ratio of \NII/\Ha=0.117 is adopted to account for the contribution of \NII\ to \Ha. These mock galaxies' \Ha\ lines are simulated with a Gaussian profile of  $\sigma$=200\,km\,s$^{-1}$ at given redshifts,  and convolved with the $H_2S1$ filter transmission curve to yield the observed \Ha+\NII\ fluxes for the mock galaxies.
 
 Similarly, we randomly assign EWs obeying the best-fitting EW distribution to the simulated galaxies of given \Ha+\NII\ fluxes and determine their $H_2S1$ and $K_{\rm s}$ magnitudes after including photon noise and sky background noises from the corresponding images. Applying the same selection criteria as presented in Figure~\ref{fig:selection}, we derive  the fraction of the selected mock galaxies in all intrinsic \Ha\ luminosity bins. Then we obtain the completeness function, as shown Figure~\ref{fig:lfcom}. As one can see that the completeness declines rapidly at  $\log (L_{\rm H\alpha}$/erg\,s$^{-1}) < 42.8$. Here the volume correction and completeness estimate are based on the redshift bin $2.225<z<2.267$,  and the final completeness curve accounts for all major effects involved in our observations and selection.  Note that the completeness curve is insensitive to the input Schechter function in our simulations and thus the determination of the intrinsic \Ha\ LF in our two overdensities is little affected by the input function in deriving detection completeness.

\begin{figure*}
\centering
\includegraphics[width=0.48\textwidth]{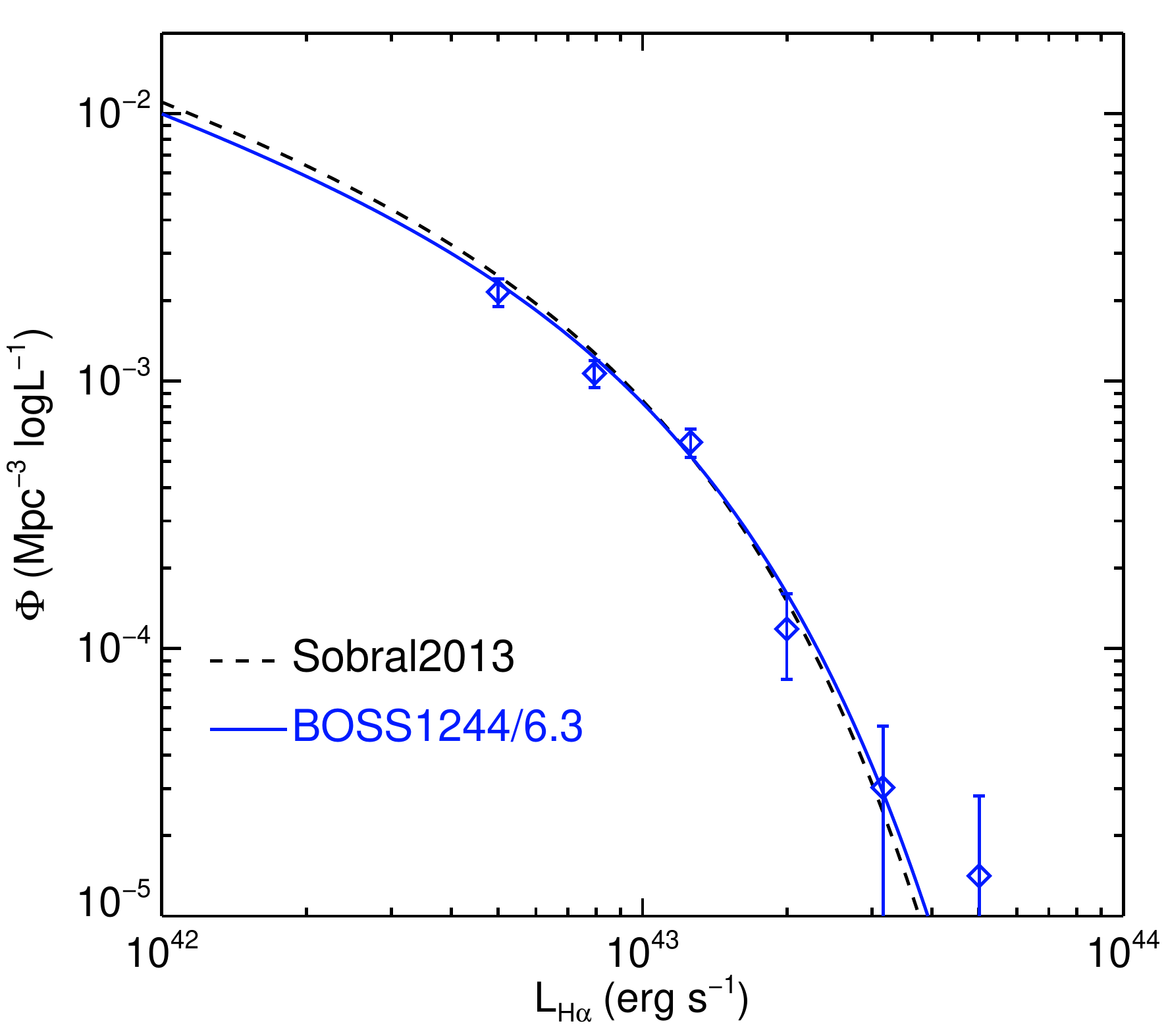}
\includegraphics[width=0.48\textwidth]{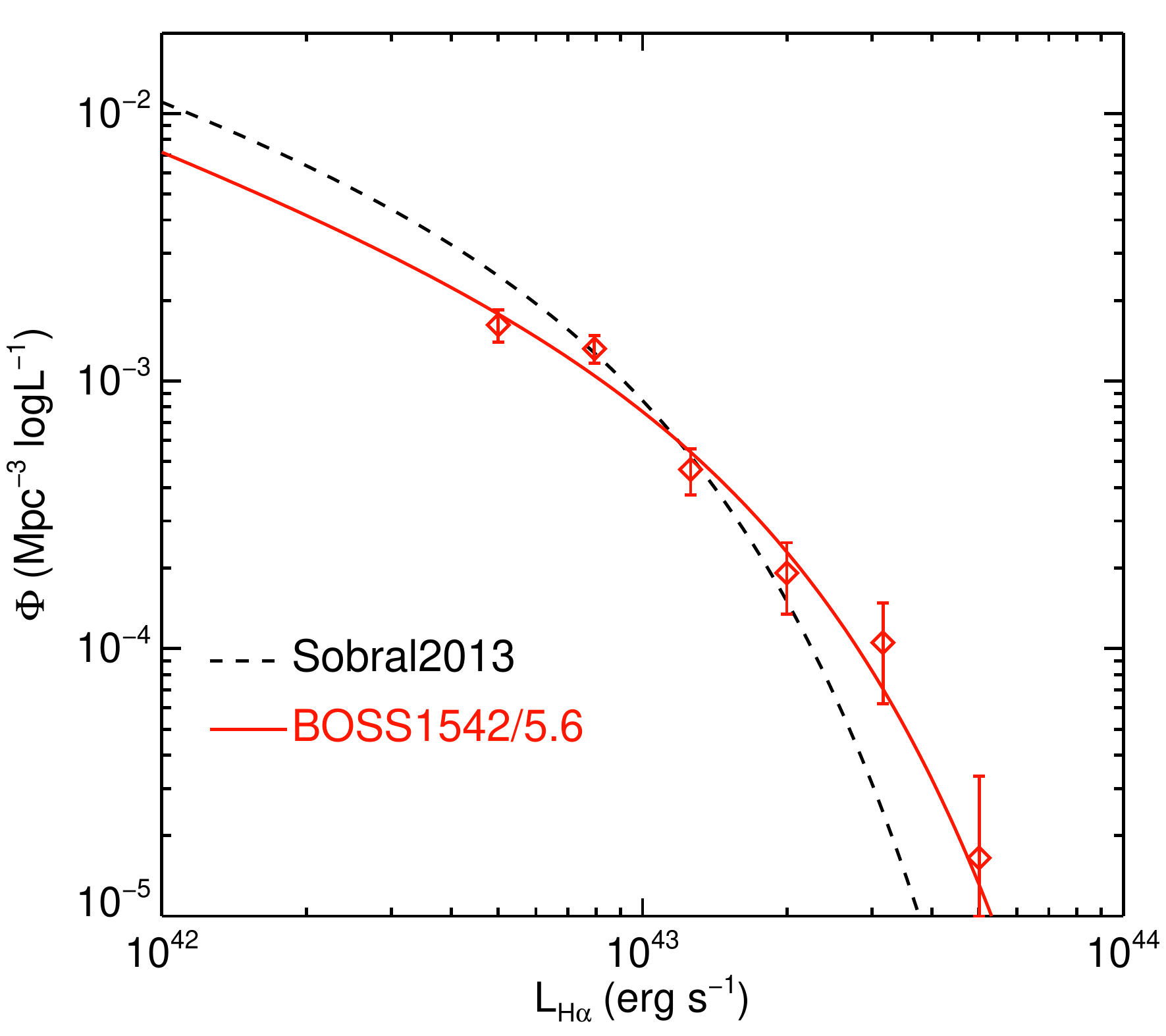}
\caption{Comparison of the \Ha\ LFs in BOSS1244 (left) and in BOSS1542 (right) with that in the general field at  $z=2.23$ from \citet{Sobral2013}.  The  \Ha\ LFs of the two overdensities are scaled down by a factor of 6.3 and 5.6, respectively. Interestingly, the \Ha\  LF of BOSS1542 shows an excess at $\log (L_{\rm H\alpha})>43.3$. This excess is not seen in BOSS1244. } \label{fig:hlf} 
\end{figure*}

\subsubsection{Determining \Ha\ luminosity function}\label{sec:dhalf}

As shown in Section~\ref{sec:iden}, We estimated 48$\pm$2 non-HAEs for both of our two emitter samples and derived that $196\pm 2$ of 244 (80\,per\,cent)  and $175\pm2$  of 223 (78\,per\,cent) emission-line objects are HAEs at $2.225<z<2.267$ belonged to the massive overdensity BOSS1244 and BOSS1542, respectively. Of these objects, five QSOs in BOSS1244 and three QSOs in BOSS1542 are excluded.  With current data and observations,  we are unable to recognize non-HAEs from the HAEs.  We subtract the non-HAEs in a statistic way when constructing \Ha\ LF.  It has been shown that the line flux distribution of the emitters in the outskirts of the two overdensities differs from that of the emitters in the dense regions, following the difference between the non-HAEs and HAEs in ECDFS. It is reasonable to draw that the outskirts contain more non-HAEs and the high-density regions are dominated by HAEs. Still, the outskirts contain a fraction of HAEs partially contributed by the dense structures, although the outskirts hold the information of true non-HAEs in the target fields.  In practice, we remove 48 emitters following the line flux distribution of the non-HAEs in ECDFS (see Figure~\ref{fig:dist}) from our samples and use the rest 143 objects in BOSS1244 and 124 objects in BOSS1542 to derive the \Ha\ LF. The difference of the line flux distribution between the non-HAEs in ECDFS and the outskirts of the two overdensities does not causes noticeable changes to the line flux distribution of the rest objects.   We point out that non-HAEs represent only $\sim$20\,per\,cent of the total emitters in the two overdensity fields. The uncertainty in estimating the number of the non-HAEs should have no significant effect on our results of the \Ha\ LFs.  The observed line fluxes of these non-HAEs tend to be relatively fainter and the vast majority ($>85$\,per\,cent) of them have line fluxes of $\log (f_{\rm line}<-15.9$. The error in correction for non-HAEs influences the faint end of the intrinsic \Ha\ LF at $\log L_{\rm H\alpha}<43.05$.

Our sample HAEs spread in $2.225<z<2.267$ over an area of 417 and 399\,arcmin$^2$,  giving a volume of 58,154 and 55,644\,$h^{-3}$\,Mpc$^3$ in BOSS1244 and BOSS1542, respectively.  We divide the sample HAEs into six \Ha\ luminosity bins over $42.6<\log (L_{\rm H\alpha})<43.8$. We calculate the volume density of HAEs at given bins after correcting for the completeness, and obtain our \Ha\ LF data points.   The Poisson noise is adopted as their errors. 
A Schechter function \citep{Schechter1976} shown below is used to fit the data points:  
\begin{equation}
\begin{split}
 & & \Phi (\log L) \,d(\log L) =\\
 & & \ln(10)\,\phistar\,10^{(\alpha+1)(\log L - \log \Lstar)}\,\exp[-10^{\log L -\log \Lstar}]\, d(\log L), 
\end{split}
\end{equation}
where $\Lstar$ refers to the characteristic luminosity, $\phistar$ is the characteristic density and $\alpha$ represents the power-law index of the faint end. The $\chi^{2}$ minimization method is utilized to determine the best-fitting parameters, giving $\log\Lstar=42.91$, $\phistar=0.0078$ and $\alpha=-1.60$ for BOSS1244, and   $\log\Lstar=43.13$, $\phistar=0.0032$ and $\alpha=-1.68$ for BOSS1542. 

We show the \Ha\ LFs of our two overdensities at  $z=2.24$ in Figure~\ref{fig:hlf}. The \Ha\ LF at  $z=2.23$ of the general field from \citet{Sobral2013} is also included for comparison.  Note that our \Ha\ LFs of BOSS1244 and BOSS1542 are scaled down by a best-matched factor of 6.3 and 5.6, respectively, consistent with  $1+\delta_{\rm gal} =6.6\pm 0.3$ and $5.9\pm 0.3$ within the uncertainties.   It is clear that  the \Ha\ LF of BOSS1244 agrees well with that of the general field, while the \Ha\ LF of BOSS1542 exhibits a prominent excess at the high end.  As can be seen from Figure~\ref{fig:hlf},  this excess is not due to an underestimate of the overdensity factor because the two data points at $\log (L_{\rm H\alpha}) < 43$ are already below the \Ha\ LF of the general field.  There are 10 and 14 objects with $\log L_{\rm H\alpha}>43.4$, accounting for 5\,per\,cent and 8\,per\,cent of HAEs in BOSS1244 and BOSS1542, respectively.  Only two objects have \Ha\ with $\log L_{\rm H\alpha}>43.6$ in each of the two overdensities.   These objects make the two data points at $\log L_{\rm H\alpha}>43.4$, and thus are critical to the high end of the \Ha\ LF.  Compared with 10 objects (5\,per\,cent of the total) in BOSS1242, the high-end of BOSS1542 consists of 14 objects (8\,per\,cent of the total), showing an excess of 50\,per\,cent for $\log L_{\rm H\alpha}>43.4$ at a 2\,$\sigma$ confidence.

We caution that our samples of HAEs possibly contain AGNs that are less luminous than quasars but  significantly contribute to \Ha\ luminosity and thus increase the high end of the \Ha\ LF of SFGs that we want to obtain.  Based on the 4\,Ms Chandra X-ray observations, \citet{An2014} identified three X-ray-detected AGNs among 56\,HAEs in the ECDFS field, being exclusively brightest HAEs with $\log(L_{\rm H\alpha})>43.5$.  This suggests an AGN fraction of 9\,per\,cent in the field when limiting HAEs to our detection depths.  The fraction of AGNs in high-$z$ protoclusters reported in previous studies is  typically several per\,cent but with large scatter, depending on the evolutionary stage, total mass and gas fraction of the protoclusters \citep[e.g.,][]{Macuga2019}.  The two bins at the high end of \Ha\ LF contain 5/8\,per\,cent of HAEs in BOSS1244/BOSS1542, comparable to the reported AGN fractions in the protoclusters. We caution that the two luminosity bins at $\log L_{\rm H\alpha}>43.4$ in our \Ha\ LFs might be seriously contaminated by AGNs.  We lack the X-ray observations to detect AGNs and get rid of them from our samples of  HAEs. 

\section{Conclusions} \label{sec:sum}

We used the WIRCam instrument mounted on CFHT to carry out  deep NIR imaging observations through narrow $H_2S1$ and broad $K_{\rm s}$ filters for identifying \Ha\ emission-line galaxies at $z=2.246\pm 0.021$ in two 20$\arcmin\times 20\arcmin$ fields, BOSS1244 and BOSS1542, where massive MAMMOTH overdensites are indicated  by  most extreme groups of IGM Ly$\alpha$ absorption systems at  $z=2.24$ over a scale of $\sim$20\,$h^{-1}$\,cMpc imprinted on the available SDSS-III spectra. The two overdensity candidates represent the extremely massive ones selected over a sky coverage of 10,000\,$\deg^2$. 

There are 244/223 emission-line objects selected with rest-frame $EW>45$\,\AA\ and $H_2S1<22.5$\,mag over an effective area of 417/399 arcmin$^2$ to the 5\,$\sigma$ depths of $H_2S1$=22.58/22.67\,mag and  $K_{\rm s}$= 23.29/23.23\,mag in BOSS1244 and BOSS1542, respectively.   Of them, $196\pm 2$ (80\,per\,cent) and $175\pm 2$ (78\,per\,cent) are estimated to be \Ha\ emitters at  $z=2.24$ in the two overdensities, in comparison with $36-40$\% of emission-line objects to be HAEs in the general field. We estimate the global overdensity factor of HAEs to be $\delta_{\rm gal}=5.6\pm 0.3$ and $4.9\pm 0.3$ in a volume of $54\times32\times32\,h^{-1}$\,cMpc$^3$ for the BOSS1244 and BOSS1542, respectively.  The overdensity factor would increase $2-3$ times if focusing on the high-density regions with a scale of $10-15$\,cMpc. The  striking excess of HAEs is convincing evidence that he two overdensities are very massive structures at $z>2$.  

The HAE density maps reveal that the two overdense structures span over 30\,$h^{-1}$\,cMpc with distinct morphologies. BOSS1244 contains two components: one low-density component connected to the other elongated high-density component.   The high-density substructure has $\delta_{\rm gal}=15$. If confirmed to be one physical structure, it would collapse into a present-day massive cluster, as suggested by simulations. In contrast, BOSS1542 manifests as a large-scale filamentary structure.  

We subtract the contribution of possible non-HAEs from our sample of HAE candidates in a statistic manner and construct \Ha\ luminosity functions for our two overdensities.  We find that the \Ha\ luminosity functions are well fit with a Schechter function. After correcting for the overdensity factor, BOSS1244's \Ha\ LF agrees well with that of the general field at the same epoch from \citet{Sobral2013}. The \Ha\ LF of BOSS1542, however, shows an excess of HAEs at the high-luminosity end at a 2\,$\sigma$ confidence. Interestingly, these HAEs with $\log L_{\rm H\alpha}>43.4$ are mostly located at the intermediate-density regions other than the density peak area. These suggest that star formation is not seriously influenced by the extremely dense environment in BOSS1244, and even plausibly enhanced in BOSS1542, although our data are unable to probe AGNs and quiescent member galaxies.  Taken together with the unbounded structures, we infer that the two  $z=2.24$ massive overdensities were undergoing a rapid assembly. 

Our results denote that the two massive overdensities at  $z=2.24$ are extremely interesting targets to  1) investigate the environment dependence of galaxy evolution; 2) address the environmental mechanisms for triggering quasar activities and address the coevolution between SMBHs and galaxies; and 3) provide constraints on  hierarchical  structure formation models and standard cosmological model.  We will address these issues in upcoming works. 

\section*{Data availability}

The data underlying this article will be shared on reasonable request to the corresponding author.

\section*{Acknowledgements}

We are grateful to the anonymous referee for helpful comments that significantly improved the manuscript. 
This work is supported by the National Key Research and Development Program of China (2017YFA0402703), the National Science Foundation of China (11773076, 11703092), and the Chinese Academy of Sciences (CAS) through a China-Chile Joint Research Fund (CCJRF \#1809) administered by the CAS South America Center for Astronomy (CASSACA). 
This research uses data obtained through the TelescopeAccess Program (TAP), which has been funded by the National Astronomical Observatories, Chinese Academy of Sciences, and the Special Fund for Astronomy from the Ministry of Finance in China.

Our observations were obtained with WIRCam, a joint project of CFHT, the Academia Sinica Institute of Astronomy and Astrophysics (ASIAA) in Taiwan, the Korea Astronomy and Space Science Institute (KASI) in Korea, Canada, France, and the Canada-France-Hawaii Telescope (CFHT) which is operated by the National Research Council (NRC) of Canada, the Institut National des Sciences de l'Univers of the Centre National de la Recherche Scientifique of France, and the University of Hawaii.












\bsp	
\label{lastpage}
\end{document}